# A Simulation-Based Framework for Leveraging Shared Autonomous Vehicles to Enhance Disaster Evacuations in Rural Regions with a Focus on Vulnerable Populations

**Alican Sevim**
**Ph.D. Student**
Department of Civil & Environmental Engineering
FAMU – FSU College of Engineering, Tallahassee, Florida, 32310
Email: asevim@fsu.edu

**Qianwen (Vivian) Guo*, Ph.D.**
**Assistant Professor**
Department of Civil & Environmental Engineering
FAMU – FSU College of Engineering, Tallahassee, Florida, 32310
Email: qguo@eng.famu.fsu.edu

**Eren Erman Ozguven*, Ph.D.**
**Associate Professor**
Department of Civil & Environmental Engineering
FAMU – FSU College of Engineering, Tallahassee, Florida, 32310
Email: eozguven@eng.famu.fsu.edu

* *Corresponding Author Email:* qguo@eng.famu.fsu.edu, and eozguven@eng.famu.fsu.edu

## ABSTRACT

Rural areas face distinct challenges during disaster evacuations, such as lower income levels, reduced risk perception, longer travel distances, and vulnerabilities of residents. Traditional evacuation methods, which often rely on state-owned buses and city-owned vans, frequently fall short of meeting the public's needs. However, rapid advancements in autonomous vehicles (AVs) are poised to revolutionize transportation and communities, including disaster evacuations, particularly through the deployment of Shared Autonomous Vehicles (SAVs). Despite the potential, the use of SAVs in rural disaster evacuations remains an underexplored area. To address this gap, this study proposes a simulation-based framework that integrates both mathematical programming and SUMO traffic simulation to deploy SAVs in pre- and post-disaster evacuations in rural areas. The framework prioritizes the needs of vulnerable groups, including individuals with disabilities, limited English proficiency, and elderly residents. Sumter County, Florida, serves as the case study due to its unique characteristics: a high concentration of vulnerable individuals and limited access to public transportation, making it one of the most transportation-insecure counties in the state. These conditions present significant challenges for evacuation planning in the region. To explore potential solutions, we conducted mass evacuation simulations by incorporating SAVs across seven scenarios. These scenarios represented varying SAV penetration levels, ranging from 20% to 100% of the vulnerable population, and were compared to a baseline scenario using only passenger cars. Additionally, we examined both pre-disaster and post-disaster conditions, accounting for infrastructure failures and road closures. According to the simulation results, higher SAV integration significantly improves traffic distribution and reduces congestion. Scenarios featuring more SAVs exhibited lower congestion peaks and more stable traffic flow. Conversely, mixed traffic environments demonstrate reduced average speeds attributable to interactions between SAVs and passenger cars, while exclusive use of SAVs results in higher speeds and more stable travel patterns. Additionally, a comparison experiment was conducted to examine whether the observed improvements with SAVs were a result of their unique capabilities or simply coincidental, by replacing SAVs with conventional buses under identical conditions.

## 1. INTRODUCTION AND LITERATURE REVIEW

Extreme weather events globally have brought to the forefront the critical necessity for innovative approaches. In 2023 alone, the US experienced a staggering 23 weather-related disasters, each causing an average loss of $1 billion, nearly tripling the annual average recorded between 1980 and 2022. These events are projected to become more frequent and severe in the future, amplifying the urgency for effective response strategies. Also, severe hurricanes, earthquakes, floods, and wildfires have had significant impacts on communities. For instance, Florida has experienced considerable damage from hurricanes. In 2018, Hurricane Michael resulted in an estimated $25 billion in damage and resulted in 74 deaths (National Oceanic and Atmospheric Administration (NOAA)). Additionally, Hurricane Irma in 2017 caused 134 deaths and approximately $77 billion in damage (Cangialosi et al., 2018). These disasters impose a financial burden and require substantial resources for recovery (Zhou et al., 2023). Both urban and rural areas face unique challenges in this context, with rural areas particularly affected by factors such as lower income levels, inadequate transportation infrastructure, and longer travel distances (Mitsova et al., 2018). Rural populations often have limited access to mobility options, reduced perception of disaster risk, and constrained budget, making them disproportionately affected by hurricanes (Miao et al., 2019). Hence, well-organized and timely evacuation plans are crucial for saving lives during disasters. While previous studies explored various evacuation strategies (Cahyanto & Pennington-Gray, 2014; Eisenman et al., 2007; Kang et al., 2007; La Greca et al., 2019; Lazo et al., 2015), conventional evacuation methods, which often depend on state-owned buses and city-owned vans for transport, may fail to encourage residents to evacuate together.

Although numerous studies have been conducted on autonomous vehicles, primarily focusing on traffic flow and safety (Copp et al., 2023; Dixit et al., 2016; Jiang et al., 2020; Kolarova & Cherchi, 2021; Rahman et al., 2021), their potential role in facilitating disaster evacuations in rural areas remains largely unexplored. Especially recent advancements in shared autonomous bus technology have gained significant attention for improving evacuation efficiency by providing flexible routing and increased capacity. Companies such as EasyMile, Auro, New Flyer, and Adastec are at the front of this technology. For instance, EasyMile's EZ10 autonomous shuttle, known for its high customizability and extensive development, exemplifies these advancements (EasyMile, 2024). Similarly, New Flyer has introduced the Xcelsior AV, North America's first heavy-duty automated transit bus, which aims to enhance safety and efficiency in public transportation (New Flyer, 2024). In addition, Adastec, in collaboration with Karsan, a Turkish commercial vehicles manufacturer, has launched Europe's first full-speed, full-size autonomous electric bus, further advancing public transport automation (Adastec, 2024). Despite this innovation, there is a notable gap in the research particularly focusing on the use of these buses in rural evacuation scenarios. To address this gap, the primary objective of this study is to develop and implement a comprehensive simulation framework for integrating Shared Autonomous Buses into emergency evacuation scenarios in a predefined rural area, prioritizing vulnerable populations such as those with disabilities, limited English proficiency, and elderly residents. According to US DOT Equitable Transportation Community (ETC) Explorer, Sumter County, Florida, a region identified as one of the most transportation-insecure regions in the state, is selected as a case study due to several factors like income, age, language barriers, and disabilities, all of which complicate effective evacuation.

The integration of autonomous vehicles into evacuation strategies shows great promise for enhancing overall evacuation efficiency. Research indicates that deploying AVs during hurricane evacuations can significantly reduce costs, network clearance times, and travel time uncertainty by optimizing large-scale evacuations through central guidance systems. Lee & Kockelman (2024) has demonstrated that evacuation efficiency improves with AVs, as they can travel with reduced headways and increase road capacity. On the other hand, for the population without vehicle access, Dynamic Ride Sharing capabilities (Fagnant & Kockelman, 2018) offer significant advantages by providing lower service times and travel costs, despite additional passenger pickups and non-direct routes. Lee & Kockelman (2024) find that larger SAV fleets may reduce waiting times but increase travel times due to more rerouting. They also highlight that an optimal configuration of 5-seat SAVs with one SAV per 14 people demonstrates effective coordination with bus schedules to manage high-demand situations and mitigate panic-induced delays. Not only optimizing evacuation costs and network clearance times but also understanding evacuation behavior and traffic demand are crucial for planning and executing effective strategies. Li et al. (2013) studied evacuation responses during Hurricane Irene in Cape May County, New Jersey, by constructing an S-shaped evacuation response curve indicating a sharp increase in evacuee departures following mandatory notices. They highlight the importance of timely and clear communication in mobilizing populations, particularly in rural areas with limited evacuation routes.

Evacuation strategies must also incorporate the optimization of evacuation routes and timing to minimize congestion and ensure efficient population movement. An integer programming formulation for optimal egress route assignment by (Stepanov & Smith, 2009) validates the effectiveness of the evacuation plan by evaluating clearance time, total traveled distance, and blocking probabilities. Furthermore, Chen & Korikanthimath (2007) emphasizes the benefits of staged evacuations over simultaneous evacuations, highlighting that appropriate staging could significantly reduce evacuation time and delays. To enhance evacuation efficiency, multi-modal evacuation strategies can benefit from incorporating both vehicular traffic and mass transit. In their study (Abdelgawad et al., 2010), they developed a framework that integrates dynamic traffic assignment for auto evacuees with optimized scheduling for transit vehicles. They found that considering in-vehicle travel time, at-origin waiting time, and fleet cost significantly reduce auto-evacuee clearance time, showing the value of mass transit in evacuation strategies. In addition, a bi-level optimization model, which minimizes both risk and travel time by determining evacuation orders and using a dynamic user equilibrium traffic model, is proposed (Apivatanagul et al., 2012). The model's application to Eastern North Carolina highlights its effectiveness in planning and evaluating evacuation strategies under different scenarios. A system-based bi-level network optimization model is also presented, aiming to balance security and stability in evacuations (Yuan et al., 2019) with residential tolerance levels and dynamic traffic assignment to address uncertainties in evacuation demand and risk distribution. Furthermore, a bi-level model is developed to optimize the location of intersections for uninterrupted flow and signal control, significantly improving evacuation efficiency (Liu & Luo, 2012).

To provide more discussions for simulation-based studies related to disaster evacuations, this review highlights research on traditional methodologies (X. Chen et al., 2012; Y. Chen et al., 2020; Lambert et al., 2013; Yazici & Ozbay, 2008), alternative evacuation proposals (Fujihara & Miwa, 2012b, 2012a, 2014; Goto et al., 2016), and select innovative studies (Ahanger et al., 2023; Iizuka et al., 2011; Na & Banerjee, 2015; Raja & Saravanan, 2022; Uno & Kashiyama, 2008; Xu et al., 2018). Due to recent severe weather events across the world, recent advancements in evacuation modeling consider the necessity for dynamic and flexible approaches that take into account both pre- and post- disaster factors along with real-time conditions during emergencies. X. Chen et al. (2012) contributed this aspect of the field by proposing a model that integrates pre- and post- disaster elements as to evaluate evacuation risks. By this approach, they intended to address a critical gap in prior research, which often focuses on pre-disaster conditions. By incorporating contingent factors such as traffic congestion and evacuee behavior, their model facilitates a more realistic and dynamic evaluation of evacuation risks. In their model application to Beijing's transportation network, they visualized risks at the neighborhood level to provide useful information for emergency planners. This study prompts a critical point of view: evacuation planning must transition from static and one-dimensional methods to adaptive and simulation-based systems capable of responding to real-time disruptions and behaviors. These insights could enhance city-level disaster preparedness, yet the challenge comes in effectively operationalizing these models within diverse urban infrastructures.

In a similar study by (Lambert et al., 2013), they reiterated the importance of a multi-layered approach to disaster evacuation in a particular emphasis of understanding transportation network performance under stress. In their methodology, they combine behavioral survey data with travel demand modeling to assess emergency evacuation scenarios in Washington, DC. They found that significant congestion and capacity strain on critical arterials would require operational adjustments such as lane reversals and strategic aid station placements. A critical point worth to mention is the impact of behavioral data on the accuracy of evacuation models, which bring further questions: how can models be refined to incorporate human decision-making under extreme disasters? Additionally, how can operational strategies be tailored to manage traffic flow while ensuring evacuees make informed, timely decisions? To follow up to these questions, the interaction of behavioral insights and demand modeling may have the key to more effective evacuation strategies in especially densely populated areas. Expanding on this, previously, Yazici & Ozbay (2008) critically examined evacuation demand models by revealing how the choice of model affects network-wide performance. In their research, the Cape May County evacuation network brings differences in performance outcomes based on the selected demand models (the S-curve, SLM, and Tweedie's approach) under constrained network capacity. This raises another question to the discussion: to what extent can these demand models be relied upon to produce trustable and/or practical predictions across varying disaster scenarios? Continuing analyzing their study, they highlight the importance of model calibration and validation in addressing real-world complexities like unpredictable human behavior, network disruptions, and cascading infrastructure failures. This discussion can be summarized to highlight the necessity for flexible, adaptable models that can be fine-tuned as conditions evolve - an essential aspect that remains overlooked in current evacuation planning efforts.

To deeper our review of evacuation methodologies, it is seen that the complexity of disaster evacuation planning necessitates more novel systems capable of optimizing routes and resources in real-time. To this end, recent studies reveal a transformative shift towards utilizing advanced computational models and technologies for evacuation strategies. To fill this aspect of gap in the literature, Uno & Kashiyama (2008) introduced a multi-agent simulation system integrated with Geographic Information Systems (GIS) to effectively model urban environments during flood scenarios. In their approach, the Dijkstra algorithm was employed to identify optimal evacuation routes, which provides a foundational framework that aligns well with the growing body of research, which focuses on personalized and context-aware evacuation systems. Visualizing the evacuee experience through virtual reality enriches understanding and informs the design of more intuitive evacuation pathways. Building on this study, Na & Banerjee (2015) proposed the Triage–Assignment–Transportation (TAT) model. In their model, they show the critical need to prioritize resources based on injury severity during large-scale disasters. By optimizing the routing of vehicles based on the available medical resources, their model stands as a good example in disaster response planning. Furthermore, the integration of multi-agent systems in their approach complements Uno and Kashiyama's framework: evacuation requires an understanding of spatial dynamics and systematic allocation of resources to maximize survivor outcomes. In this context, further innovations emerge from Iizuka et al. (2011), who present a disaster evacuation assist system utilizing ad-hoc networks. By applying a Distributed Constraint Optimization Problem (DCOP) algorithm, their system estimates evacuee locations to optimize evacuation timing and demonstrate the potential for peer-to-peer communications in mitigating congestion. This approach was followed by another study from Ahanger et al. (2023), who proposes an IoT-inspired evacuation framework that apply real-time data collection through smart devices. In their framework, they employ IoT to collect real-time data, edge computing to detect emergencies using support vector machines (SVM), and cloud computing to optimize evacuation paths. Their experimental simulations show significant improvements in decision-making efficiency, energy efficiency, and reliability, with metrics such as a 5.23-second temporal delay, 96.69% F-measure, and 4.56 mJ energy consumption. Furthermore, "CLOTHO", an IoT-based evacuation planning system, as detailed by Xu et al. (2018), advances this context by introducing a mobile cloud computing platform that employs an Artificial Potential Field (APF) algorithm for crowd evacuation planning. CLOTHO integrates a mobile cloud computing platform for data collection and analytics, utilizing an APF algorithm as its core. The system's focus on calming evacuees and addressing psychological factors is highlighted as a critical advantage in enhancing overall evacuation efficiency. Finally, Raja & Saravanan (2022) introduces a Multi-Agent Deep Reinforcement Dijkstra (MADRD) algorithm designed to optimize the navigation of Connected Autonomous Vehicles (CAVs) in large-scale disaster evacuations. Their algorithm operates within a 6G-assisted environment, allowing CAVs to learn from each other's experiences to determine optimal routes in the presence of unpredictable moving obstacles. In their model results, the framework reduced fuel consumption by 40%, increased road throughput by 45%, and lowered total evacuation time by 34% compared to existing state-of-the-art methods. All together with these studies, that were featured in this review or not, they make a strong case for incorporating a range of technologies- from GIS and IoT to machine learning and CAVs- to create effective disaster evacuation systems.

Additionally, we believe that, as shown in the literature, there is significant potential for further exploration and research into the use of AV technologies for evacuation purposes. Traditional evacuation methods that rely exclusively on passenger cars require and/or assume all evacuees to own a vehicle, which brings challenges for those who depend on public transport and may leave them at a disadvantage during emergencies. Furthermore, elderly and/or disabled residents with vehicles may also find it difficult to drive evacuations, exacerbating their vulnerability. Additionally, limited English proficiency among some evacuees could hinder their ability to follow evacuation instructions properly. In contrast, while on-demand services in general offer benefits like flexibility and dynamic routing, SAVs present additional advantages that make them especially promising for vulnerable populations. Beyond the typical features of on-demand services, SAVs offer enhanced accessibility through features such as ramps, spacious interiors, and securement systems for wheelchairs, ensuring that individuals with mobility challenges can easily access and use these services. Furthermore, SAVs incorporate advanced technological capabilities, including door-to-door service, dynamic route optimization, and remote operation. These features not only improve accessibility but also contribute to reducing crash rates and alleviating traffic congestion.

Following this, the study seeks to contribute to the literature through the following research questions and contributions:

1. *How can SAVs be integrated into emergency evacuation plans to address the specific needs of vulnerable populations in rural areas?*
   By utilizing SUMO traffic simulation capabilities, this research develops a comprehensive framework to assess the potential of SAVs in rural disaster evacuation scenarios by considering both pre and post disaster conditions. The focus on rural areas, which are often overlooked in transportation research, addresses a critical gap in current evacuation planning, particularly for regions with sparse transportation networks and high concentrations of vulnerable populations.
2. *What are the benefits and limitations of using SAVs compared to traditional passenger cars in rural evacuation scenarios, particularly with regard to traffic flow, congestion reduction and accessibility?*
   This study goes beyond general on-demand transportation solutions by analyzing the advanced capabilities of SAVs, such as dynamic routing. These functionalities are assessed to determine their effectiveness in improving traffic conditions, minimizing congestion, and enhancing overall accessibility for evacuees.
3. *What insights can be gained from simulated evacuation scenarios to inform the broader adoption of SAV technologies for disaster response?*
   Unlike traditional studies, this research prioritizes the unique needs of vulnerable populations, such as those with mobility challenges, limited English proficiency, and elderly residents. By emphasizing inclusivity, the study provides actionable insights for designing equitable evacuation strategies. Through detailed simulation of evacuation scenarios, it also generates empirical evidence on the feasibility and applicability of SAVs in rural contexts. In the near future, these findings will have the potential to inform policymakers, emergency planners, and technology developers about the potential of SAVs for disaster management.

It is also essential to recognize that SAV technology remains in its early stages, and the study is limited to simulations. There is a considerable way to go before SAVs achieve the necessary technological readiness, regulatory approval, and societal acceptance to be widely adopted for evacuation purposes (Dia & Javanshour, 2017; Miller et al., 2022; Moody et al., 2020; Paddeu et al., 2020).

The subsequent sections of this paper are organized as follows: first, the experimental setup and methodology of the study will be outlined, including key assumptions, the configuration of the simulation environment, road network generation, network arrangements, vehicle configurations, microscale modeling and behavior, and the configuration and implementation of population demand for evacuation scenarios.

This will be followed by a discussion of the simulation results, along with an analysis of the study's limitations and conclusions.

## 2. EXPERIMENTAL SETUP & METHODOLOGY

Refer to **Figure 1** for an animated representation of the study's conceptual framework. This visualization illustrates how the evacuation process incorporating SAVs could be structured and executed in real-life scenarios to provide a context before presenting methodological details.

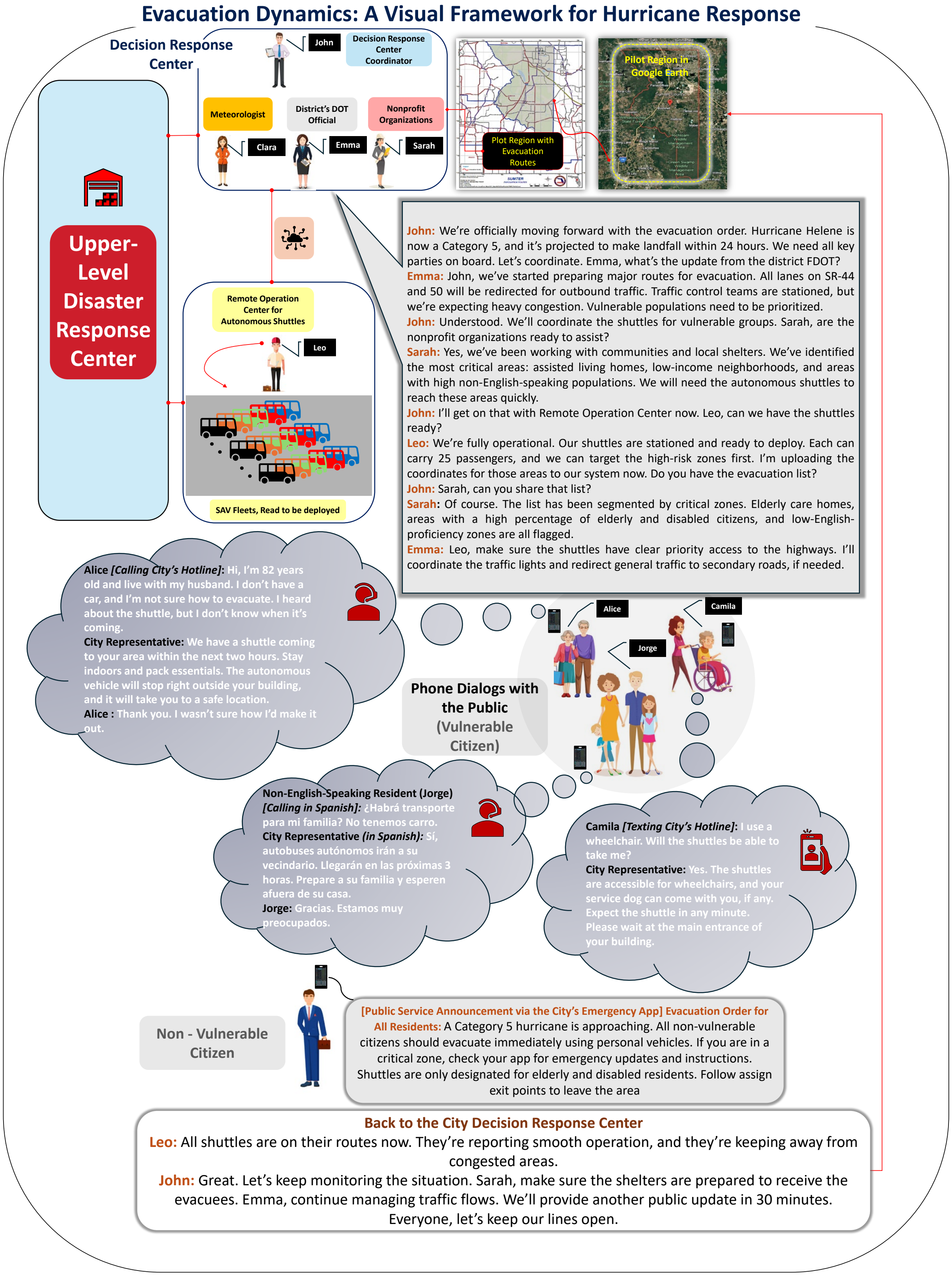

**Figure 1.** An Illustrated Animation of the Proposed Evacuation Process, Highlighting Coordination among City Officials, Nonprofits, and Vulnerable Populations in Response to Hurricane Helene

### 2.1. Assumptions and Simulation Setup for Disaster Evacuation in Sumter County, Florida

Given the diverse nature of natural disasters and the unique characteristics of each region, it is essential to establish clear assumptions for the simulation setup and population characteristics. In addition, disasters impact populations differently, but their effects are particularly severe on vulnerable populations. Thus, selecting a representative region for a vulnerable population is critical. For this study, Sumter County is selected as it is notable for its predominantly elderly population. According to Census Data (U.S. Census Bureau, 2024), its population is 137,265, with 6,836 individuals aged 85 years and older. Additionally, 27,938 individuals have disabilities, and approximately 2,000 residents have low levels of English proficiency. Notably, according to USDOT's ETC explorer (USDOT ETC Explorer, 2024) around 27,000 people in the area in Disadvantaged Census Tracts, which is nearly 30% of the population. Furthermore, the overall disadvantage component score for transportation insecurity in the region is 85% (indicating a high cumulative burden of transportation-related disadvantages based on factors such as access, health, and vulnerability metrics). These characteristics make the selected region an ideal representation of a vulnerable population for the simulation. On the other hand, in terms of vehicle configurations, two types of transportation modes—SAVs and regular passenger cars—are adopted. Notably, each SAV is designed to accommodate an average of 25 vulnerable individuals, exclusively serving this population to provide a practical unit of evacuation capacity. To determine SAV occupancy capacity, different autonomous shuttle companies are reviewed, and an average value is selected to reflect real cases. Meanwhile, one passenger car is allocated for every five individuals who do not fall into the vulnerable population category to distinguish the efficiency of SAVs in emergency situations. SAVs are assumed to be immediately available upon passenger request (Dia & Javanshour, 2017). Given that the study area is predominantly residential and car-dependent, we assume that individuals not assigned to an SAV rely exclusively on their personal vehicles for evacuation. Consequently, large vehicles such as trucks and buses were excluded from the main simulation setup, except for conventional buses included in a comparative analysis (see Section 3.3), as they represent only a small fraction of the total vehicle population. Furthermore, this study examines hurricane evacuation scenarios under two distinct conditions: (1) a fully operational road network without closures and (2) a compromised network with road closures due to emerging infrastructure failures caused by early flooding. In the first case, it is assumed that the hurricane has not yet made landfall, and the road network remains intact. This scenario represents an evacuation initiated 1-2 days before the hurricane's expected arrival to provide evacuees with adequate time to reach safety. The operational state of the road network in this scenario represents an ideal pre-disaster condition where traffic flow is uninhibited by physical disruptions or flooding. In contrast, the second case models a more challenging scenario where the hurricane's approach begins to impact on the road network. Roads susceptible to flooding and structural failure are identified based on FEMA's National Flood Hazard Maps. It is important to note that while the primary focus of this study is pre- and post-disaster evacuation, the framework is designed to be adaptable to other emergency contexts. Also, the simulation does not explicitly model background traffic within the county or along exit roads. Instead, a simplified approach assumes varying proportions of the total vehicle count may implicitly represent such traffic. For traffic along exit roads, as long as the evacuation flow is not entirely obstructed, this simplification is considered reasonable. External traffic jams on exit routes outside the county's jurisdiction, which could create bottlenecks, are beyond the scope of this study. It is also assumed that upon the issuance of an evacuation order, the entire population will begin evacuating simultaneously, resulting in a mass evacuation with no exceptions.

### 2.2. Simulation Environment Setup

An open-source traffic simulator is utilized, Simulation of Urban Mobility (SUMO). SUMO is a detailed, multi-modal traffic flow simulation platform that operates in both spatially continuous and temporally discrete modes (Krajzewicz et al., 2012). In SUMO (**Figure 2**), several key elements are considered to configure the evacuation traffic: the road network, including nodes and unidirectional edges representing streets, waterways, tracks, bike lanes, and walkways, with each edge having a specific geometry described by a series of line segments and consisting of one or more lanes running in parallel; the road infrastructure, including elements such as traffic lights and other infrastructure components; and the population demand, which involves creating the traffic demand along with vehicle configurations and additional files (routes, bus stops, and etc.) to be used in SUMO's configuration files (Lopez et al., 2018). With all these elements together, the methodology for handling and configuring of each component will be first detailed.

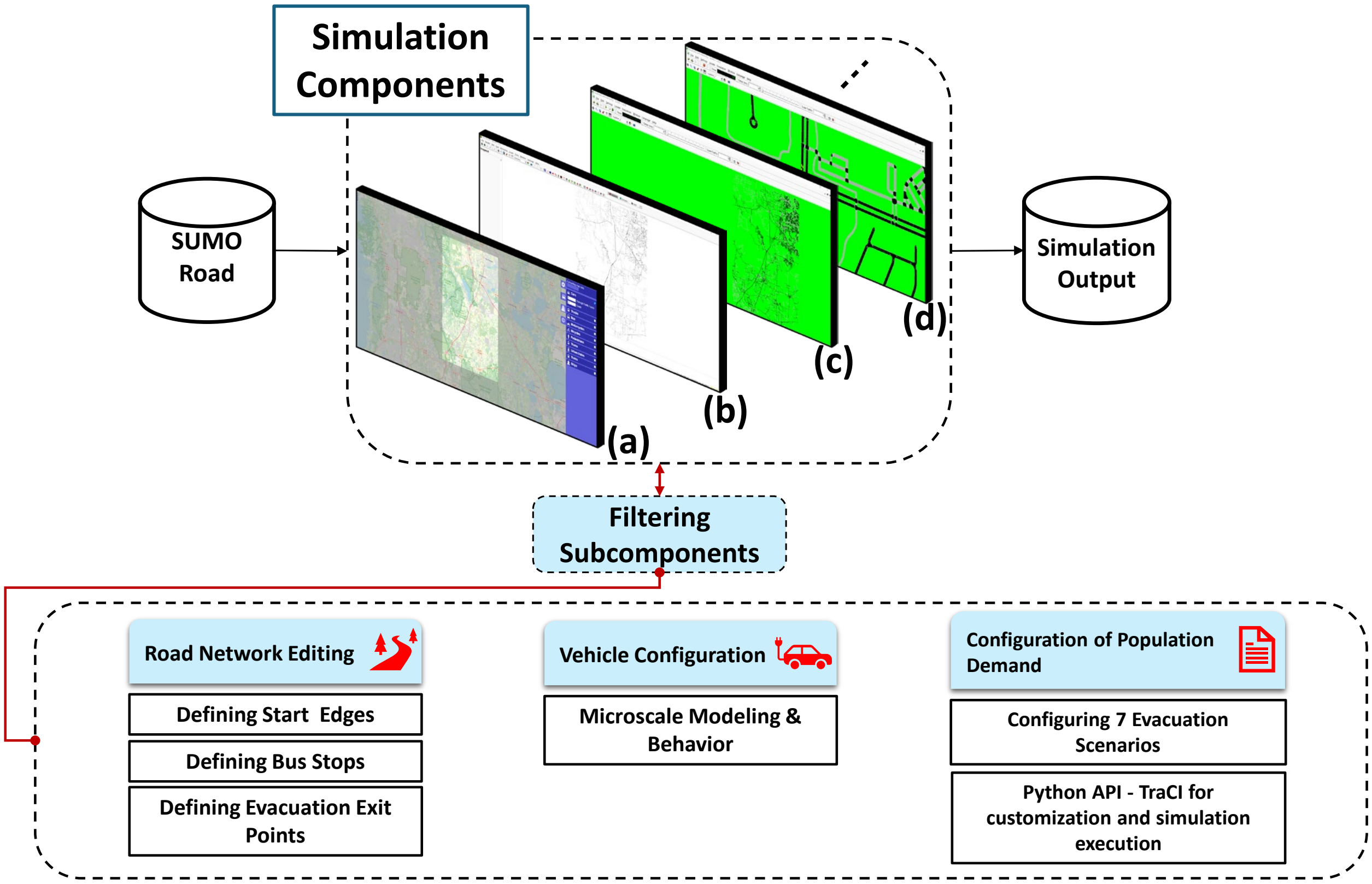

**Figure 2.** Simulation Environment Setup Flow: (a) Visualization of the map excerpt obtained from OpenStreetMap Wizard. (b) Extracted map displayed on the Netedit interface. (c, d) Representation of the map within the SUMO interface

*2.2.1. Road Network Generation*

We obtain open-source road network data from OpenStreetMap (OSM), which provides detailed geographic features encoded as XML files. The map selection process involves specifying the area of interest using latitude and longitude ranges (SUMO Documentation, OSMWebWizard Tutorial, 2024). (**Figure 2a**). The map data contains elements such as nodes, edges (roads), relations, and associated attributes that define the physical and functional network properties. Roads, including curved segments, are represented as a series of connected line segments, while intersections are defined by nodes. While the OSM Web Wizard automates much of the conversion process, we further performed manual inspection and validation to address potential inaccuracies. These include incorrect lane connections, missing or misclassified road attributes (e.g., priority, speed limits, or road types), and inconsistencies in traffic controls like stop signs and traffic lights. For better network accuracy, the inspection process begins with arterial roads and progresses to local and residential roads. Arterial roads are initially inspected for their type, priority, and connectivity, followed by a more detailed inspections of lane conditions and traffic control elements. During this phase, misconfigurations – such as incorrect road connections or missing attributes – are identified and resolved.

### *2.2.2. Network Editing Using Netedit*

Netedit (**Figure 2b**), a network editing tool within SUMO, is utilized to incorporate critical network elements such as bus stops, evacuation exit routes, and the evacuation paths for SAVs and passenger cars. SUMO's main network elements include edges, lanes, junctions (also referred to as nodes or intersections), and connections. The selected road network consists of 27,148 nodes and 66,546 edges. Additionally, network elements such as bus stops are managed in separate auxiliary files, which are embedded into the configuration file and loaded at the start of each simulation. In terms of operating features, Netedit operates in three major editing modes: network-related objects, traffic-related objects, and data objects. These modes facilitate the inspection, deletion, and addition of various objects, with some modes being common across all supermodes and others specific to particular (SUMO Documentation, Netedit, 2024). The first step in the process involves identifying evacuation routes and assigning SAVs to designated routes, which include bus stops and exit points. Here, the aim is to prevent bottlenecks and improve evacuation efficiency during simulation execution. Pick-up locations are determined based on population density and vulnerability, targeting areas with higher percentages of elderly, disabled, or low-income residents with limited access to private vehicles. The key considerations for bus stop placement are listed below:

- The proximity of clustered bus stops significantly influences overall utilization and operational efficiency; therefore, the bus stops should be spaced appropriately throughout the region,
- Simultaneous bus arrivals at clustered stops should be avoided to minimize potential delays and enhance passenger flow,
- Bus stops should be positioned strategically within approximately 500 feet of residences to maximize convenience for residents, while also maintaining a minimum distance of 800-1000 feet between adjacent stops to prevent overcrowding,
- Accessibility to bus stops must be ensured for all passengers without causing congestion, which could hinder the efficiency of the transportation network and negatively impact the overall simulation results.

After identifying pick-up locations, bus stops are added using Netedit's "Additional" mode. Several experiments are conducted to observe the network and determine the optimal locations for bus stops. The final configuration is saved in a separate file named *additionals.add.xml*. Following this, specific start edges and exit points are defined to facilitate their evacuation from the region. These start edges and exit points are selected to represent realistic evacuation routes and to distribute the traffic evenly across the network. For regular passenger vehicles, their origins were randomly selected, typically located in high-density residential regions. These origins were primarily concentrated on the upper east side of the region. While shelter locations are considered in other studies, such as (Dulebenets et al., 2020), this work focuses on large-scale, county-level evacuations under severe disaster scenarios. As a result, localized evacuations of small regions and shelter-in-place strategies were not considered. Furthermore, shelters were assumed to be unavailable or inadequate to provide sufficient protection during disasters, which necessitates a full-scale evacuation (Y. Chen et al., 2020; Xie & Turnquist, 2011). To simulate these conditions, evacuation exit points were defined based on the region's official evacuation plans, which includes arterial routes. These exit points (**Figure 3**) include key routes such as SR 44, I-75, FL Turnpike, SR 50, CR 48, CR 476, FL 471, and SR 35 or US 301.

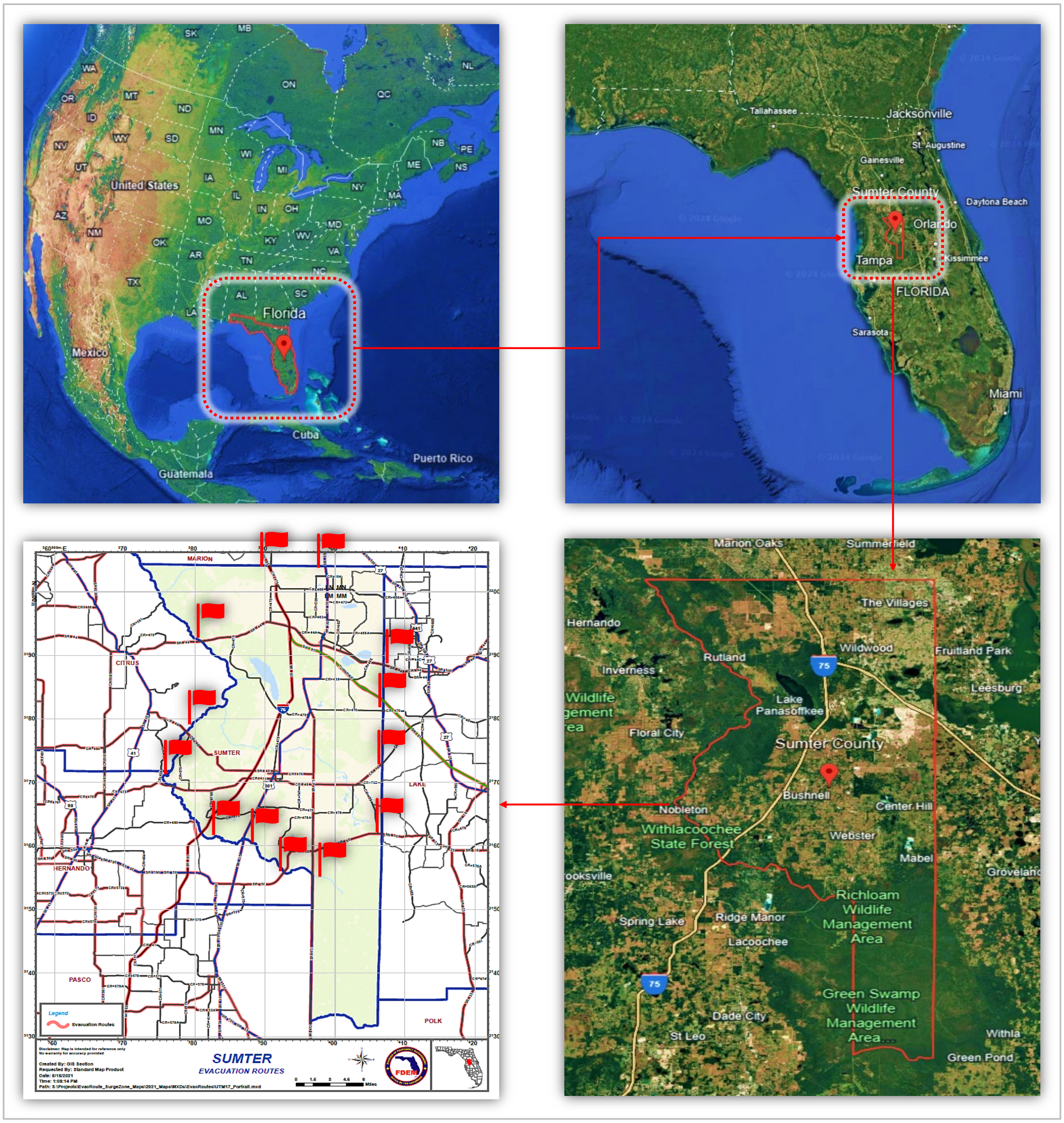

**Figure 3.** Map of the Target Region Showing Highlighted Evacuation Routes and Selected Exit Points

### *2.2.3. Vehicle Configurations*

In the simulation setup, two primary vehicle types are modeled: SAVs and Human-Driven Vehicles (HDVs). **Table 1** presents vehicle configurations, which illustrate the differences in dynamic characteristics and operational behaviors.

*2.2.3.1 SAVs*

SAVs are configured to simulate fully autonomous buses with enhanced safety and operational efficiency. In this study, SAVs are modeled at the microscale level following the approach outlined in the work by (Bamdad Mehrabani et al., 2023). In their study, they presented a multiclass simulation-based dynamic traffic assignment model for mixed traffic flows of connected and autonomous vehicles and human-driven vehicles. Building on this research, we applied distinct parameter sets for two different scenarios: pre-disaster evacuation and post-disaster evacuation conditions. For the pre-disaster evacuation scenario, SAVs operate under relatively relaxed traffic conditions, which depicts a near-steady state of traffic flow prior to disaster-induced disruptions. Each SAV is equipped with a rerouting device, and they are configured with initiating route recalculations every 60 seconds. This interval, combined with a pre-period of 300 seconds and an adaptation interval of 60 seconds, enables continuous recalibration of vehicle trajectories in response to incremental variations in traffic density or flow. The pre-period parameter prevents premature rerouting during the initial phase of the simulation. By adding this parameter to the vehicle configuration, we ensure that vehicles stabilize within the network before dynamic rerouting begins. On the other hand, during the post-disaster evacuation scenario, the dynamic rerouting parameters are recalibrated to address the heightened complexity of traffic dynamics caused by road closures, congestion, and safety-critical constraints. To this end, the rerouting period is extended to 180 seconds to balance computation intensity by reducing frequent rerouting due to partially compromised road network. The pre-period remains fixed at 300 seconds, while the adaptation interval is also extended to 180 seconds for reflecting a deliberate dampening of hyperreactive rerouting behaviors to prevent oscillatory traffic patterns. Additionally, we introduced a *device.rerouting.threshold* parameter with a value of 0.1. With this parameter, we make sure that SAVs only update their paths if the new route provides at least a 10% improvement in travel time. This modification adds another layer of constraint to reduce unnecessary rerouting and stabilizes SAV behavior, even in dynamic traffic conditions. To further mimic the evacuation conditions, operational speed for SAVs adhere to official road segment speed limits during pre-disaster scenarios to compliance with standard traffic regulations. However, during post-disaster conditions, SAVs adopt a reduced speed profile derived from a speed distribution constrained under the speed limit via speed factor parameter of 0.9. The specific vehicle parameters for SAVs remain consistent across both scenarios and include an acceleration rate of 3.5 m/s², a deceleration rate of 4.5 m/s², and an emergency deceleration rate of 8 m/s². They also maintain a minimum gap of 1.5 meters, with vehicle dimensions of 8 meters in length, 2.5 meters in width, and 3.4 meters in height.

*2.2.3.2 HDVs*

HDVs, representing regular passenger cars, have an acceleration rate of 2.6 m/s², a deceleration rate of 4.5 m/s², and an emergency deceleration rate of 8 m/s². These vehicles have a minimum gap of 2.5 meters between vehicles. Also, the dimensions of HDVs are a length of 5 meters, a width of 1.8 meters, and a height of 1.5 meters, with a capacity to carry up to 5 passengers. To realistically simulate driving behavior, under pre-disaster conditions, the speed of HDVs is dynamically restricted according to the official speed limits of their respective road segments. A Gaussian distribution is applied to determine each vehicle's maximum speed for a realistic variability of typical driving behavior. In contrast, during post-disaster scenarios, a speed factor parameter is applied to reduce HDVs' operational speeds. This reduction accounts for the degraded roadway conditions and increased safety requirements for a more controlled speed range.

*2.2.3.3 Microscale Modeling and Behavior*

The movement of both SAVs and HDVs is simulated using advanced car-following and lane-changing models (Lopez et al., 2018) at the microscale level in SUMO. The longitudinal dynamics of these vehicles are governed by the Krauss car-following model (Krauß, 1998). For SAVs, this model is modified to reflect their enhanced capabilities, providing faster and safer behavior compared to HDVs. By this medication, to ensure that SAVs, equipped with full automation (Level 5), collisions can be avoided (sigma parameter is set to zero, see the **Table 1**) by reacting within the acceleration bounds of both the leading and following vehicles. This approach is consistent with the methodology outlined in (Bamdad Mehrabani et al., 2023), which effectively simulates mixed traffic flows of connected autonomous vehicles (CAVs) and HDVs. In terms of lateral dynamics, the LC2013 lane-changing model is employed (Lopez et al., 2018). A critical parameter in this configuration is *lcAssertive*, which quantifies a vehicle's propensity to accept smaller gaps during lane changes. For SAVs, the *lcAssertive* value is set to 0.7, indicating a more conservative lane-changing strategy while HDVs have a higher *lcAssertive* value of 1.3, reflecting a more aggressive approach to lane changes. By these distinctions, we underscore SAVs' optimized, cautious maneuvers compared to the more selfish behavior of HDVs.

**Table 1.** Specifications for the Car-Following Model in SUMO

| | **Mingap (*m*)** | **Acceleration (*m/sn*$^2$)** | **Deceleration (*m/sn*$^2$)** | **Emergency Deceleration (*m/sn*$^2$)** | **Sigma** | **IcAssertive** |
|---|---|---|---|---|---|---|
| **SAV Shuttle** | 1.5 | 3.5 | 4.5 | 8 | 0 | 0.7 |
| **HDV** | 2.5 | 2.6 | 4.5 | 8 | 0.5 | 1.3 |

Mingap: The distance to the vehicle ahead when stopped in traffic *(meters)*
Accel: The rate at which vehicles can accelerate *(meters per second squared)*
Decel: The rate at which vehicles can decelerate *(meters per second squared)*
Emergency Decel: The maximum deceleration capacity of vehicles during emergencies *(meters per second squared)*
Sigma: The measure of driver imperfection, ranging from 0 to 1
IcAssertive: Quantifying a vehicle's propensity to accept smaller gaps during lane changes

*2.2.4. Configuration and Implementation of Population Demand for SUMO Evacuation Scenarios*

In **Table 2**, the notations used have been listed.

**Table 2.** List of Notations

| **Indices** | **Definition** |
|---|---|
| $P$ | Total population requiring evacuation |
| $P_{sub}$ | Subpopulation (e.g., Elderly > 85, With Disabilities) |
| $C_{PV}$ | Capacity of a passenger vehicle |
| $C_{SAV}$ | Capacity of an SAV |
| $N_{PV}$ | Total number of passenger vehicles required |
| $N_{SAV}$ | Total number of SAVs required |
| $E$ | Number of starting edges (for passenger vehicles) |
| $B$ | Number of bus stops (for SAVs) |
| $X$ | Number of exit points (fixed at 13) |
| $R$ | Number of remaining vehicles after equal distribution |
| $T_w$ | Total evacuation time window |
| $t_{dep}$ | Departure time for vehicles in the population demand file |
| $\lambda_{sav}(t)$ | Time-varying dispatch rate representing the number of SAVs |
| $\lambda_{pv}(t)$ | Time-varying dispatch rate representing the number of PVs |
| $f_{sav}(i, j, t)$ | The flow of SAVs between origin $i$ and destination $j$ at time $t$ |
| $f_{pv}(i, j, t)$ | The flow of PVs (Passenger Vehicle) between origin $i$ and destination $j$ at time $t$ |
| $R_{SAV}$ | Denoting the remaining vehicles (SAVs) after the initial assignment to bus stops |
| $R_{PV}$ | Denoting the remaining vehicles (PVs) after the initial assignment to starting edges |
| $\boldsymbol{F}_{SAV}\ (i, j)$ | Vehicle flow (SAVs) between bus stop $i$ and exit point $j$ |
| $\boldsymbol{F}_{PV}\ (i, j)$ | Vehicle flow (PVs) between starting edge $i$ and exit point $j$ |

To conduct the SUMO simulations for evacuation scenarios, a comprehensive setup is configured, including a configuration file encompassing additional files, OpenStreetMap files, and crucially, population demand files (SUMO, Population Demand, 2024). Population demand files are vital in accurately modeling how traffic participants select routes through the road network to reach their destinations. The focus is on prioritizing the vulnerable population in Sumter County, utilizing population metrics from the Census database. Initially, the total population and specific vulnerable groups: elderly individuals aged 85 and above, people with disabilities, and those with limited English proficiency, are identified. Although the majority of the region's residents are native English speakers, a substantial number still face language barriers, which could potentially hinder their ability to follow evacuation instructions effectively.

In this study, a bi-level model has been adopted to simulate the integration of SAVs into disaster evacuation scenarios. Notably, the upper level represents the decision-making authority, where the objective is to optimize the dispatch rates of SAVs and passenger vehicles within a constrained evacuation time window $T_w^*$. In the upper level, the decision-making process involves selecting the optimal time window for evacuation and ensuring that all vehicles are dispatched efficiently within this time frame, mimicking the scenarios where everyone follows the evacuation orders. At the lower level, the problem has been formulated with two distinct methodologies: a User Equilibrium (UE) model for passenger vehicles and a dynamic routing model for SAVs. For PVs, the UE model ensures that each individual user selfishly selects their route to minimize personal travel costs based on current traffic conditions by following a principle, where no driver can reduce their travel cost by unilaterally changing routes. Here, the objective is to balance the flow across the network such that all routes reach an equilibrium based personal optimization. On the other hand, SAVs are modeled using a dynamic routing approach. Contrary to PVs, which adhere to static route choices, SAVs continuously adapt to real-time traffic conditions. This dynamic routing is formulated as a dynamic optimization problem, where SAVs periodically update their routes to reduce congestion and ensure a more stable traffic distribution. By this bi-level structure, we ensure that the interaction between the system's optimization (e.g., dispatching rates and evacuation timeframes) and individual user behavior (e.g., routing choices based on travel costs) is fully captured. Upper-level optimizations influence user behavior by constraining the evacuation window and vehicle dispatch rates, while the lower-level UE formulation captures how users react to these conditions by choosing the best routes based on travel costs.

A stepwise approach is used in the simulation setup, followed by no fleet size restrictions for SAVs. SAVs are assigned based on population metrics with a total SAV occupancy of 25, under the assumption that these on-demand shuttles are immediately available to the target population. To enhance realism, as mentioned, an upper-level restriction is incorporated into the simulation, reflecting the official evacuation time window. After experimenting with various durations, a 6-hour window is determined to be optimal. A 12-hour timeframe has been proved excessively long, failing to yield realistic traffic density, while shorter durations overloaded the system. The 6-hour window is a strategic choice to ensure that all vehicles in the population demand file had departures times within this timeframe. To more elaborate on this rationale, let $T_w$ represent the total evacuation time window in hours. In the experimentation, as said, the optimal window for realistic density is chosen as 6 hours, denoted as $T_w^* = 6$. An evacuation window of 12 hours ($T_w = 12$) is found to be excessively long, whereas a window shorter than 6 hours leads to a system overload. To incorporate these into the model, the objective function of the upper level can be defined:

$$T_w^* = \arg\min_{T_w}\left(\mathfrak{L}(T_w)\right), \tag{1}$$

where $\mathfrak{L}(T_w)$ is a loss function defining the trade-off between traffic density and evacuation efficiency. The goal is to minimize the loss $\mathfrak{L}(T_w)$, which increases if the window is either too long or too short. For instance, if $T_w < T_w^*$ the system challenges with overload due to the insufficient time for vehicles to depart, while $T_w > T_w^*$ leads to unrealistic traffic density and inefficiency. Hence, the optimal evacuation window is the result of balancing these competing factors. To this end, let $t_{dep}$ denote the departure time for vehicles in SUMO demand file. To ensure that all vehicles are dispatched within the designated evacuation time window $T_w^*$, the departure times for both SAVs and PVs must satisfy the constraint as below:

$$0 \leq t_{dep} \leq T_w^*. \tag{2}$$

In terms of dispatch rate formulation, the dispatch rate functions $\lambda_{sav}(t)$ and $\lambda_{pv}(t)$, are defined for SAVs and PVs, respectively, representing the number of vehicles dispatched per unit time $t$. The total dispatch rate must guarantee that all necessary vehicles are dispatched within the predefined time window:

$$\int_0^{T_w^*} \lambda_{sav}(t)\, dt = N_{SAV}, \tag{3}$$

$$\int_0^{T_w^*} \lambda_{pv}(t) dt = N_{PV}. \tag{4}$$

The dispatch rates are time-varying and adjusted based on real-time traffic conditions. For SAVs, which are equipped with rerouting devices, the dispatch rate $\lambda_{sav}(t)$ is dynamically adjusted every 60 or 180 seconds based on the evacuation scenario. For PVs, the dispatch rate $\lambda_{pv}(t)$ is adjusted using a time-adaptive factor to avoid the traffic bottlenecks and system overloads. In addition, both SAVs and PVs must satisfy the flow conservation law across the network. This states that the total number of vehicles dispatched over time matches the number of required vehicles by maintaining demand satisfaction at both origin and destination points while minimizing traffic congestion. To formulize this, let $f_{sav}(i, j, t)$ and $f_{pv}(i, j, t)$ represent the flow of SAVs and PVs between origin $i$ and destination $j$ at time $t$. The total flow must meet the vehicle demand:

$$\sum_{i,j} \int_0^{T_w^*} f_{sav}(i, j, t) dt = N_{SAV}, \tag{5}$$

$$\sum_{i,j} \int_0^{T_w^*} f_{pv}(i, j, t)\, dt = N_{PV}. \tag{6}$$

Lastly, the total dispatch rate for both SAVs and PVs at any given time $t$ is defined by the superposition of the individual dispatch rates:

$$\lambda_{total}(t) = \lambda_{sav}(t) + \lambda_{pv}(t). \quad (7)$$

Furthermore, by the following algorithm snipped, it is demonstrated that how the departure times for the population demand in line with upper-level restrictions have been adjusted. In the analysis, evacuation schedules are adjusted to align with the designated time spans by modifying the start and end times of HDV flows.

**Algorithm: Adjusting Departure Times for Demand Based on Upper-Level Restrictions**

```
def adjust_departure_times (base_schedule, time_span, demand_data):
    if not demand_data or 'flows' not in demand_data:
        raise ValueError ("Invalid demand data format.")

    hdv_flows = [flow for flow in demand_data ['flows'] if flow ['type'] == 'hdv']
    if not hdv_flows:
        raise ValueError ("No HDV flows found in the dataset.")

    adjustment_log = []

    for idx, flow in enumerate (hdv_flows):
        t_begin = base_schedule + idx * time_span
        t_end = t_begin + time_span
        flow ['begin'] = t_begin
        flow ['end'] = t_end
        adjustment_log.append ({"flow_id": flow['id'], "new_begin": t_begin, "new_end": t_end})

    return demand_data, adjustment_log
```

Here, we perform a dual-phase recalibration of evacuation schedules for pre- and post- disaster scenarios with temporal flow adjustments. Initially, pre-disaster evacuation leverages a linear mapping $T_i = t_0 + i \cdot \Delta t$, where $T_i$ denotes adjusted departure times for HDV flows indexed by $i$, with $t_0$ as the base schedule and $\Delta t$ defined by a fixed span constraint. Using XML parsing, the algorithm identifies $F_{HDV} \subset F$, the subset of all flows satisfying type = '*hdv*', and iteratively adjusts $[T_i^{begin}, T_i^{end}]$ by explicitly updating their attributes. This approach is similarly extended to SAV flows. For post-disaster evacuation, departure times are calibrated using a Gaussian-derived S-curve model. Further details on the post-disaster S-curve calibration are provided in Section 3.2.

Following this, seven evacuation scenarios are examined in the study for both pre and post disaster evacuations (**Table 3**), starting with a baseline scenario and progressing to the seventh scenario. The reason for choosing seven scenarios is to systematically analyze the impact of gradually integrating SAVs on evacuation efficiency. In each scenario, a certain percentage of the vulnerable population is evacuated by SAVs, with the remaining population assigned to regular passenger cars. Notably, in the baseline scenario, the entire population (totally 137,265 individuals) is assigned to passenger cars, each with a capacity of five occupants. After that, all passenger vehicles are distributed evenly across 150 starting edges, which are identified based on the population-density with designated evacuation start points on the map. As mentioned, evacuation endpoints are based on the region's evacuation plan, with flows distributed evenly across 13 fixed exit points. Note that passenger cars followed a basic user equilibrium (UE) traffic assignment, where each user selfishly chose the most convenient path based on the lowest travel cost. From the second to the sixth scenario, SAVs are gradually integrated into the evacuation process by considering 20%, 40%, 60%, 80%, and 100% of the vulnerable population, respectively. For instance, in the second scenario, only 20% of the vulnerable population is assigned to SAVs, while the remaining individuals and the general population relied on their personal passenger vehicles. By this strategy, it is reflected the incremental availability of SAVs, starting with a limited number and increasing their presence in subsequent scenarios. In addition, note that when fewer than 100 SAVs are required, the count is rounded up to 100 to maintain alignment with the predefined 100 bus stops and ensure at least one SAV departs from each stop for vulnerable populations. For instance, in Scenario 2, where 16 SAVs are needed for the low English proficiency (LEP) category, assigning each SAV to a unique bus stop could result in no SAVs departing from some bus stops for this category. In order to address this, at least one SAV departs from each bus stop for the LEP category, regardless of the actual number of SAVs needed. Although ensuring to keep at least one SAV for each bus stop when actual demand is less than 100 could have introduced extra traffic congestion and excessive use of resources, any excessive traffic congestion has not been seen during the experiment. However, further assignment strategies will be considered going forward to further optimize the distribution. Moreover, the demand distribution for SAVs and HDVs across bus stops and starting edges is configured to maintain balanced traffic flow and efficient evacuation dynamics. For example, in the first scenario, the distribution of 183 passenger vehicles for every 149 starting edges and 186 vehicles departing from one edge ensures even traffic distribution during the evacuation. A similar strategy was followed for the rest of the scenarios, and, in the seventh scenario, the entire population is evacuated using SAVs, significantly contributing to reducing traffic congestion. Demand distribution methodology is discussed further in the following section.

#### *2.2.4.1 Trip Distribution*

To construct the evacuation trip distribution, a framework has been introduced below that extends beyond the simple allocation strategies. This formulation utilizes matrix multiplications to ensure efficient and balanced distribution of both SAVs and PVs in the network across starting edges and bus stops, while guaranteeing optimal flow towards designated exit points. With this approach, it has been provided a robust and scalable solution that accommodates complex evacuation scenarios with variable constraints and vehicle capacities.

To begin, let the total population $P$ be decomposed into $n$ distinct subpopulation $P_{sub}$, where $i \in \{1, 2, \ldots, n\}$. The total number of vehicles for each subpopulation $i$, for both SAVs and PVs, can be formulated as:

$$N_{SAV_i} = \left\lceil \frac{P_{sub_i}}{C_{SAV}} \right\rceil, \tag{8}$$

$$N_{PV_i} = \left\lceil \frac{P_{sub_i}}{C_{PV}} \right\rceil, \tag{9}$$

where $N_{SAV_i}$ and $N_{PV_i}$ are column vectors, which represent the vehicle demand for each subpopulation across starting edges $E$ or bus stops $B$, and $C_{SAV}$, $C_{PV}$ represent the respective capacities of SAVs and PVs. In addition, the ceiling operator ensures that the number of vehicles is rounded up to the nearest whole number.

In order to distribute vehicles across the starting edges and bus stops, it has been defined two allocation matrices $A_{PV} \in \mathbb{R}^{E \times n}$ and $A_{SAV} \in \mathbb{R}^{B \times n}$, where:

$$A_{PV}(i,j) = \left\lfloor \frac{N_{PV_i}(j)}{E} \right\rfloor, \tag{10}$$

$$A_{SAV}(i,j) = \left\lfloor \frac{N_{SAV_i}(j)}{B} \right\rfloor. \tag{11}$$

These matrices ensure the vehicles have been distributed evenly across the available starting edges and bus stops. For the remainders of the division, $R_{PV}$ and $R_{SAV}$, denote the remaining vehicles:

$$R_{PV}(i) = N_{PV_i}(j) \bmod E, \tag{12}$$

$$R_{SAV}(i) = N_{SAV_i}(j) \bmod B. \tag{13}$$

The term "*mod*" refers to modulo operation. Specifically, for each subpopulation (e.g., of SAVs or PVs), the modulus operation is used to determine how many vehicles remain after evenly distributing them across the starting edges or bus stops. The remainder vehicles $R_{PV}$ and $R_{SAV}$ are then assigned iteratively to the first $R$ starting edges or bus stops, respectively:

$$A_{PV}(i, 1 : R_{PV}) = A_{PV}(i, 1 : R_{PV}) + 1, \tag{14}$$

$$A_{SAV}(i, 1 : R_{SAV}) = A_{SAV}(i, 1 : R_{SAV}) + 1. \tag{15}$$

To allocate the vehicles across exit points, let $M_{PV} \in \mathbb{R}^{E \times X}$ and $M_{SAV} \in \mathbb{R}^{B \times X}$ represent Origin-Destination (OD) matrices, where $X$ is the number of exit points from the target region, which is given from **Table 2**. The number of vehicles routed from each starting edge or bus stop to each exit point is given by:

$$M_{PV} = A_{PV} D_{PV}, \tag{16}$$

$$M_{SAV} = A_{SAV} D_{SAV}, \tag{17}$$

where $D_{PV} \in \mathbb{R}^{n \times X}$ and $D_{SAV} \in \mathbb{R}^{n \times X}$ denote the diagonal matrices representing the proportion of vehicles distributed to each exit point for each subpopulation. These matrices have been constructed to ensure that vehicles are distributed evenly across the exit points while minimizing congestion. To handle the remainders in exit point allocation, the remaining vehicles $R_{M_{PV}}$ and $R_{M_{SAV}}$ are iteratively distributed across exit points $X$, following a round-robin approach:

$$M_{PV}(i,j) = M_{PV}(i,j) + 1 \quad \text{for } j = 1, 2, \dots, R_{M_{PV}}, \tag{18}$$

$$M_{SAV}(i,j) = M_{SAV}(i,j) + 1 \quad \text{for } j = 1, 2, \ldots, R_{M_{SAV}}. \tag{19}$$

For example, the passenger vehicle OD matrix $M_{PV}$ has dimensions of $(E, X)$, representing the allocation $E$ starting edges to $X$ exit points. Similarly, the OD matrix for SAVs, $M_{SAV}$, has dimensions of $(B, X)$:

$$M_{PV} = \begin{pmatrix} M_{11} & M_{12} & \cdots & M_{1X} \\ M_{21} & M_{22} & \cdots & M_{2X} \\ \vdots & \vdots & \ddots & \vdots \\ M_{E1} & M_{E2} & \cdots & M_{EX} \end{pmatrix}, \ M_{SAV} = \begin{pmatrix} M_{11} & M_{12} & \cdots & M_{1X} \\ M_{21} & M_{22} & \cdots & M_{2X} \\ \vdots & \vdots & \ddots & \vdots \\ M_{B1} & M_{B2} & \cdots & M_{BX} \end{pmatrix}. \tag{20}$$

In regard to the overall optimization of vehicle flows can be expressed as minimizing a traffic cost function $C(M_{PV}, M_{SAV})$, representing travel times across the road network. Thus, the objective is to minimize total evacuation time and at the same time to ensure a balanced distribution across the exit points:

$$min(M_{PV}, M_{SAV}) = \sum_{i=1}^{E}\sum_{j=1}^{X} T_{ij} M_{PV}(i,j) + \sum_{i=1}^{B}\sum_{j=1}^{X} T_{ij} M_{SAV}(i,j), \tag{21}$$

subject to:

$$\sum_{j=1}^{X} M_{PV}(i,j) = A_{PV}(i), \tag{22}$$

$$\sum_{j=1}^{X} M_{SAV}(i,j) = A_{SAV}(i), \tag{23}$$

where $T_{ij}$ represents the travel time between starting edge/bus stop $i$ and exit point $j$. The constraints ensure that the total number of vehicles at each starting edge and bus stop matches the OD matrix distributions.

When the OD matrices $M_{PV}$ and $M_{SAV}$ are constructed and balanced, they have been applied to the traffic network by iterating over each exit point $J \in X$. The vehicle flows from each starting edge and bus stop are then mapped to the road network based on the dynamic routing for SAVs achieved through rerouting or user equilibrium traffic assignment (UE) for regular passenger vehicles:

$$\boldsymbol{F}_{PV}(i,j) = M_{PV}(i,j) \cdot \boldsymbol{SP}_{ij}, \tag{24}$$

$$\boldsymbol{F}_{SAV}(i,j) = M_{SAV}(i,j) \cdot \boldsymbol{SP}_{ij}, \tag{25}$$

where $\boldsymbol{SP}_{ij}$ is the shortest path matrix, denoting the optimal paths between starting points and exit points. By this, it has been ensured that vehicles follow the most efficient routes by minimizing overall evacuation time while distributing the traffic evenly through the network.

*2.2.4.2 User Equilibrium for PVs*

In transportation networks, once each individual driver selects their route from an origin to a destination, the equilibrium condition is met when no driver can reduce their travel time by unilaterally switching routes. This has been referred to as User Equilibrium (UE). The principle behind this phenomenon is that all available routes between an origin-destination pair experience the same travel time due to each driver choosing the least time-consuming path. Once equilibrium is reached, any changes to the selected route will not result in shorter travel times.

According to (Sheffi, 1985), the User Equilibrium problem is formulated mathematically as an optimization problem that minimizes a specific objective function and at the same time satisfies constraints on flow conservation and non-negativity.

*UE Objective Function:*

Notably, let the road network be represented as directed graph $G = (V, A)$, where $V$ is the set of nodes or intersections and $A$ is the set of directed links or road segments. For each link $(i, j) \in A$, following variables are defined:

- $x_{ij}$ denotes the follow-on road segment $(i, j)$,
- $t_{ij}\ (x_{ij})$ is the travel time on link $(i, j)$, which is a function of the flow $x_{ij}$.

Thus, the objective function is the minimization of the sum of the integrals of the link travel time functions over all road segments:

$$\min \sum_{(i,j) \in A} \int_0^{x_{ij}^{pv}} t_{ij}^{pv}(\omega) d\omega. \tag{26}$$

This formulation provides the total travel cost across the network, where the integral of the latency function $t_{ij}\ (\omega)$ captures the cumulative travel time for all drivers using road segment $(i, j)$.

*Flow Conservation Constraints:*

For each OD pair $c \in C$, where $C$ is the set of OD pairs, let:

- $y_k^c$ represents the flow on path $k$ which connects origin-destination pair $c$,
- $K_c$ is to be set of available paths for OD pair $c$.

  The flow on link $(i, j)$ must be equal to the sum of flows on all paths passing through link $(i, j)$,

which is weighted by an incidence matrix $a_{ij}^k$, where $a_{ij}^k = 1$ if path $k$ uses link $(i, j)$, and 0 otherwise. Hence, the arc flow conservation constraints can be expressed as:

$$x_{ij} = \sum_{c \in C} \sum_{k \in K_c} a_{ij}^k y_k^c \qquad \forall (i, j) \in A \tag{27}$$

In this equation, the total flow on link $(i, j)$ is the sum of the path flows from all OD pairs that use that link.

*Demand Satisfaction Constraints:*

The total flow on each OD pair must satisfy the demand $d_c$ between the origin and destination. Thus, the demand satisfaction constraint for each OD pair $c \in C$ is given by:

$$d_c = \sum_{k \in K_c} y_k^c \qquad \forall_c \in C \tag{28}$$

Through this constraint, the total flow on all paths connecting an OD pair equals the total demand for that pair.

*Non-negativity Constraints:*
Lastly, the flow on each road segment and path must be non-negative:

$$x_{ij} \geq 0 \quad \forall (i,j) \in A, \quad y_k^c \geq 0 \quad \forall_c \in C, \ \forall k \in K_c \tag{29}$$

By this constraint, it has been guaranteed that flows are physically feasible, i.e., there are no negative vehicle flows.

*2.2.4.3 Dynamic Routing for SAVs*

To achieve a more realistic representation of vehicle behavior, particularly the autonomous decision-making capabilities of SAVs, a dynamic routing framework was employed. We utilized SUMO's in-built dynamic rerouting functionality for SAVs to adaptively determine their routes based on evolving traffic conditions. The approach equips SAVs with the capability to periodically update their optimal paths by considering both the current traffic state and historical trends, thus accounting for network disruptions such as road closures and traffic bottlenecks. To this end,
Let:

- $r_t^{sav}(i,j)$ represent the rerouting probability for an SAV on link $(i,j)$ at time $t$,
- $p_t^{sav}(i,j)$ be the proportion of SAVs using link $(i,j)$ at time $t$.

The rerouting capability of SAVs can be formulated as a dynamic optimization problem where the SAVs adjust their paths based on the current state of traffic. The updated flow for SAVs on link $(i,j)$ at time $t + \Delta t$ is given by:

$$x_{ij}^{sav}(t + \Delta t) = r_t^{sav}(i,j) \cdot x_{ij}^{sav}(t) + \left(1 - r_t^{sav}(i,j)\right) \cdot p_t^{sav}(i,j). \tag{30}$$

With this dynamic strategy, SAVs continuously adapt their routes based on real-time traffic conditions. Additionally, they are configured to minimize the total travel time across the network by periodically rerouting. Therefore, the dynamic traffic assignment for SAVs can be expressed as:

$$\min_{r_t^{sav}, p_t^{sav}} \sum_{(i,j) \in A} \int_0^{x_{ij}^{sav}} t_{ij}^{sav}(\omega) d\omega, \tag{31}$$

subject to:

$$x_{ij}^{sav}(t + \Delta t) \geq 0 \quad \forall (i,j) \in A, \quad p_t^{sav}(i,j) \geq 0 \quad \forall (i,j) \in A. \tag{32}$$

The core of this methodology integrates Dijkstra's algorithm (Sniedovich, 2006) for shortest-path computation at dynamically applied regular intervals. At each recalibration step, the algorithm considers a weighted graph representation of the road network, where edges are dynamically assigned weights corresponding to the average travel times of the respective road segments. These travel times are not obtained solely from instantaneous traffic data, which may result in myopic decisions, but instead are computed as a moving average of historical travel times over a defined temporal window. The smoothing mechanism reduces volatility caused by transient congestion spikes, resulting in more stable and globally efficient routing behavior. Mathematically, the travel time for a given road segment $e_{ij}$ at time $t$ is computed as follows:

$$\tau_{ij}(t) = \frac{1}{N}\sum_{k=1}^{N} \tau_{ij}^{k}, \tag{33}$$

where $\tau_{ij}^{k}$ represents the observed travel time during the $k$-th monitoring period, and $N$ is the number of periods included in the moving average. This aggregated measure is used to define the edge weight $\omega_{ij}(t) = \tau_{ij}(t)$ in the graph model. Using the weighted graph, each SAV executes Dijkstra's algorithm to identify the shortest path to its destination at intervals of $\Delta t$ by staying responsive to changes in the network. The periodicity $\Delta t$ is calibrated to balance computational usage and real-time adaptability. The SAV's routing strategy further incorporates mechanisms to discourage shortsighted, locally optimal routing choices that might exacerbate congestion at critical network nodes. This is achieved by integrating penalties for edge with high variance in travel times over the historical window, represented as:

$$\sigma_{ij}^{2} = \frac{1}{N}\sum_{k=1}^{N} (\tau_{ij}^{k} - \tau_{ij}(t))^{2}. \tag{34}$$

Edges with high $\sigma_{ij}^{2}$ values are dynamically penalized to reduce their likelihood of being included in the shortest-path calculation and balance demand across the network and mitigate cascading congestion effects.

Finally, by combining the User Equilibrium model for passenger vehicles and the Dynamic Routing model for SAVs, a comprehensive traffic assignment formulation has been obtained. The overall objective is to minimize the total travel time throughout the network for both vehicle types:

$$\min\left(\sum_{(i,j)\in A}\int_{0}^{x_{ij}^{pv}} t_{ij}^{pv}(\omega)d\omega + \sum_{(i,j)\in A}\int_{0}^{x_{ij}^{sav}} t_{ij}^{sav}(\omega)d\omega\right), \tag{35}$$

subject to the flow conservation and demand satisfaction constraints for both vehicle classes.

After setting up the demand and configuration files for all predefined scenarios, *sumo-gui* is utilized with the Python API named TraCI (Traffic Control Interface) for customization and simulation execution. SUMO simulated each vehicle separately by employing mesoscopic network flow behaviors to reduce simulation time for large networks compared to detailed microscopic models. This setup ensured a systematic and structured approach to population demand configuration for evaluating evacuation scenarios, incorporating both SAVs and HDVs.

**Table 3.** Population Demand Distributions Across Different Scenarios - From the second to the sixth scenario, we incrementally integrated SAVs, covering 20% to 100% of the vulnerable population. For instance, only 20% of this group used SAVs in the second scenario, with the rest using their personal vehicles.

| Defined Scenarios | Population Category | Vulnerable Population | Assigned Transportation Mode | Aggregated Number of SAVs and PVs |
|---|---|---|---|---|
| **1st Scenario *(Baseline)*** | Elderly > 85 | 6836 | PV | 27,453 (100%) |
| | With Disability | 27,938 | | |
| | LEP *(Low English Proficiency)* | 2000 | | |
| | Remaining Population | 100,491 | | |
| **2nd Scenario *(20%)*** | Elderly > 85 | 1367 | SAV | 424 (1.6%) |
| | With Disability | 5588 | | |
| | LEP | 400 | | |
| | Remaining Population | 29,419 + 100,491 | PV | 25,982 (98.4%) |
| **3rd Scenario *(40%)*** | Elderly > 85 | 2734 | SAV | 657 (2.6%) |
| | With Disability | 11,175 | | |
| | LEP | 800 | | |
| | Remaining Population | 22,065 + 100,491 | PV | 24,512 (97.4%) |
| **4th Scenario *(60%)*** | Elderly > 85 | 4102 | SAV | 936 (3.9%) |
| | With Disability | 16763 | | |
| | LEP | 1200 | | |
| | Remaining Population | 14,709 + 100,491 | PV | 23,040 (96.1%) |
| **5th Scenario *(80%)*** | Elderly > 85 | 5469 | SAV | 1,214 (5.32%) |
| | With Disability | 22,351 | | |
| | LEP | 1600 | | |
| | Remaining Population | 7354 + 100,491 | PV | 21,569 (94.68%) |
| **6th Scenario *(100%)*** | Elderly > 85 | 6836 | SAV | 1,492 (6.9%) |
| | With Disability | 27,938 | | |
| | LEP | 200 | | |
| | Remaining Population | 100,491 | PV | 20,100 (93.1%) |
| **7th Scenario (SAV-only)** | Elderly > 85 | 6836 | SAV | 5,491 (100%) |
| | With Disability | 27,938 | | |
| | LEP | 200 | | |
| | Remaining Population | 100,491 | | |

## 3. SIMULATION RESULTS

### 3.1. Pre-Disaster Evacuation Scenario Modeling

In this section, comparative performance of various evacuation scenarios (pre-disaster) is discussed by highlighting the impacts of incorporating SAVs. Through the analysis, the key evacuation metrics collected during simulation experiments such as total travel time of the vehicles during evacuation, average distance traveled, traffic volume, average speed, and congestion index across seven different evacuation scenarios are examined. It should be noted that as discussed earlier, the seventh scenario represents the complete evacuation of the regional population to designated safe points using SAVs. However, given this technology is still in its early stages and considering the current fleet size limitations, this scenario should be considered as an extreme case in this study.

In **Figure 4a**, it is observed that mean travel time in SAV involved scenarios, which represents the average travel time per vehicle at any given moment, is generally less than the baseline scenario, where evacuees are solely rely on their personal vehicles. As more SAVs are integrated into the evacuation process, it becomes evident that total evacuation time at any given moment is typically shorter than in the baseline scenario, as highlighted in **Figure 4b** (yellow highlighted region). While the second and third scenarios do not show significant differences from the baseline scenario, the reduction in travel time becomes more evident after the fourth scenario.

Moreover, the total evacuation duration, represented on the x-axis, clearly demonstrates that incorporating more SAVs into the evacuation process helps to reduce overall evacuation time span and results in quicker clearance from the region. Considering the vulnerability of the region, shorter evacuation time span is particularly advantageous for vulnerable communities, as it directly impacts their in-vehicle time.

The differences among scenarios in terms of both total travel time and evacuation duration were attributed to the rerouting capability of SAVs. As mentioned, these vehicles are configured to adjust their routes every 60 seconds to find the most optimal path to the exit points. Although this feature may initially result in increased travel time, especially during the early stages of the evacuation (as highlighted in **Figure 4a** in orange), it effectively alleviates congestion and reduces the evacuation time span by distributing the evacuation traffic more effectively, as indicated in **Figure 4a** (red-highlighted region). Therefore, while integrating SAVs may lead to longer travel times in the early stages of the evacuation, it also helps to mitigate longer durations.

In **Figure 4,** the baseline scenario is marked by the black dashed lines and the seventh scenario is marked by the red dashed lines. As shown in **Figure 4a** and **Figure 4b**, all scenarios, except for the seventh, show a similar increasing trend in travel time during the initial stages of evacuation, likely reflecting the onset of congestion. The most notable improvement is observed in the seventh scenario, where the entire population evacuates using SAVs. While this may indicate a highly efficient evacuation with reduced evacuation time in managing evacuation traffic more efficiently compared to scenarios relying heavily on passenger cars, as indicated earlier, this scenario should be approached with caution.

Furthermore, the variability in travel times across scenarios can be linked to the differing levels of traffic congestion, influenced by the proportion of SAVs. The reduction in the overall evacuation time span with higher integration of SAVs suggests that they not only enhance traffic flow but also optimize route selection more effectively than traditional passenger cars.

**Figure 4c** illustrates the average distance traveled across different evacuation scenarios. In the initial phase, all scenarios exhibit a gradual increase in average distance which indicates the initial movement of vehicles from their starting points. Notably, the baseline scenario, along with scenarios two through six, indicates moderate variability and a steady rise, reflecting traffic dynamics dominated by passenger cars. On the other hand, the seventh scenario, where 100% of the population uses SAVs, demonstrates a sharp increase in the initial phase of the evacuation. Furthermore, the seventh scenario along with the other SAV involved scenarios exhibit the most significant deviation compared to the baseline scenario, with a sharp increase in average distance highlighted as light blue. This spike would highlight the capability of SAVs to manage large-scale evacuations effectively by quickly dispersing traffic and minimizing congestions but at the same time, this may lead to more distance traveled to find the most optimum path in especially later stages of the evacuation. In addition, it is observed that notable anomalies include sudden drops in some scenarios towards the end of the evacuation period, which is likely due to the rapid clearance of vehicles as they reach evacuation endpoints, which is also observable in the rest of the metrics. Furthermore, it is generally observed that, except for the seventh scenario, the average distance traveled in scenarios involving SAVs is slightly higher than in the baseline scenario, especially in the final stage of the evacuation (highlighted in light blue). This might suggest that although SAVs might result in reduced travel times, they often cover more distance likely due to their effective rerouting capabilities. However, when the entire population relies on SAVs, as in the seventh scenario, the average distance traveled follows a distinct pattern compared to the baseline scenario.

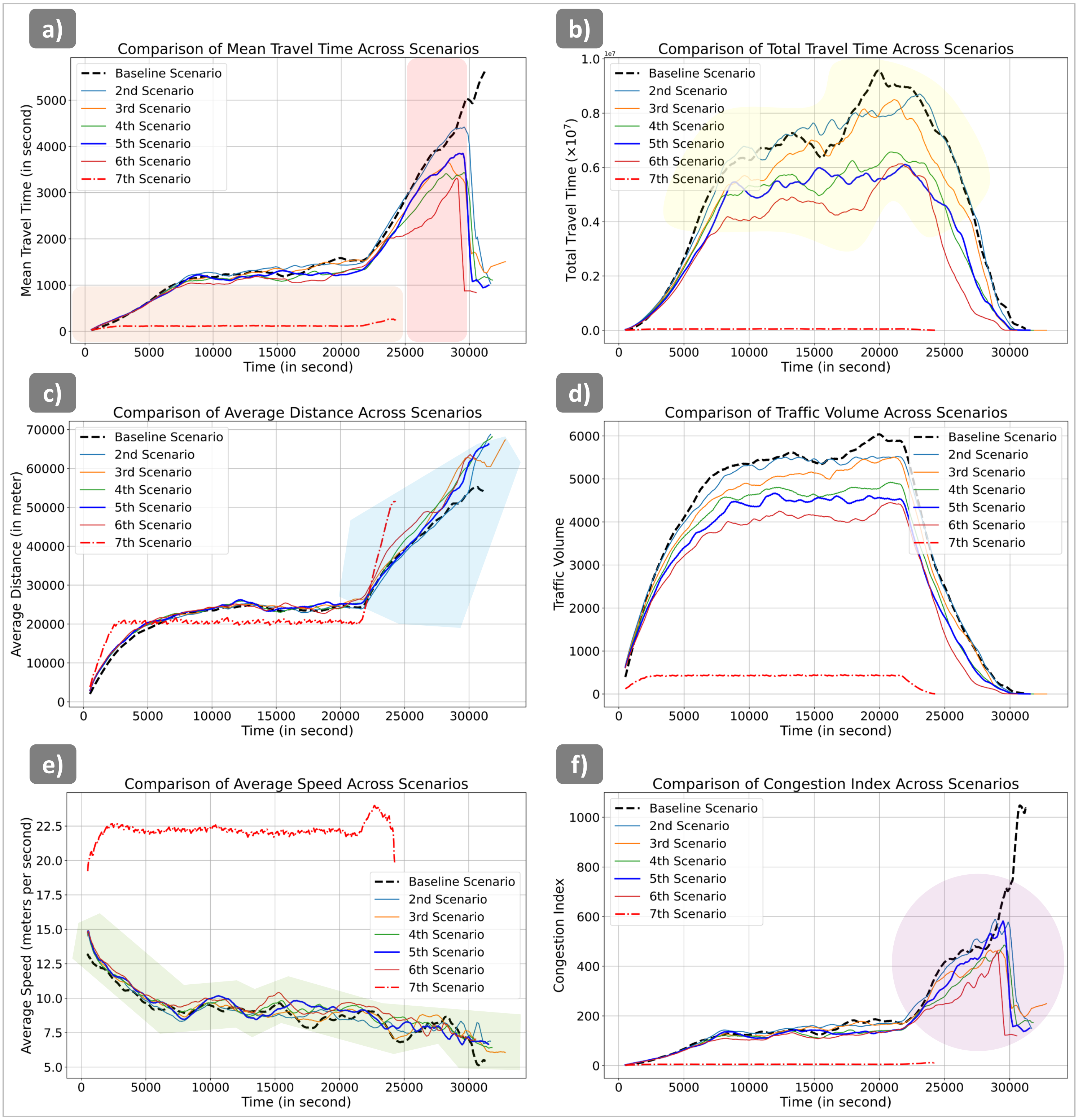

**Figure 4.** Comparison of Pre-Disaster Evacuation Metrics Obtained from SUMO Experiments Across Seven Distinct Scenarios

Thus, integrating SAVs into evacuation scenarios results in different average distances traveled compared to the baseline scenario. These variations indicate that the average distance traveled should be considered when evaluating the efficiency of evacuation management, as it can influence both traffic flow and overall evacuation outcomes.

**Figure 4d** presents the traffic volume within the network at any time given moment across different scenarios. The baseline scenario, along with scenarios two through six, shows a steady increase in traffic volume, peaking around the mid-phase of the evacuation. This is followed by a period of sustained high traffic volume before a rapid decline in the final stage. In the initial phase, scenarios two through six closely follow the baseline scenarios, indicating minimal impact from this level of SAV integration. However, after the mid-point of the evacuation, distinct patterns begin to emerge, with these scenarios exhibiting reduced peak volumes and more stable traffic flows. This may suggest that higher levels of SAV integration contribute to more efficient traffic management. The peaks in these scenarios are less evident, and the traffic volume curves are more uniform, likely reflecting the positive impact of SAVs on traffic distribution.

In contrast, the seventh scenario shows a distinct pattern reaching its peak volume earlier than the other scenarios and sustains a steady volume throughout the evacuation, rather than following the characteristic curve observed in the other cases. This suggests a more rapid mobilization of SAVs and a more continuous and steady flow of traffic, potentially resulting from the higher operational efficiency of SAVs in managing large-scale evacuations.

It is also important to note that the traffic volume represented in the graph does not reflect the total number of vehicles used in the evacuation but rather the total number of vehicles present in the network at any given time. This is due to the simulation environment, where we introduced the vehicles in batches, and some vehicles leave the network as they complete their evacuation routes. Therefore, the graph primarily illustrates the dynamic traffic volume within the network over time.

In regard to the speed metric, in **Figure 4e**, all scenarios initially exhibit a high average speed, which is likely due to the beginning of the evacuation with relatively free-flowing traffic. In mixed traffic environments, represented by scenarios two through six, the overall average speed fluctuates compared to the baseline scenario—sometimes higher, sometimes lower—depending on the specific moment (highlighted in green). This outcome is expected, as the integration of SAVs into the traffic influences the flow, which requires passenger cars to adjust their trajectories accordingly. SAVs, which we configured to prioritize efficient travel patterns rather than selfish driving, contribute to a reduction in average speed within these mixed scenarios. This decrease is likely due to the more coordinated and optimized routing behaviors of SAVs, which may slow down the overall flow but lead to a more organized and distributed traffic pattern by reducing bottlenecks.

Conversely, for scenarios exclusively involving SAVs (i.e., without mixed traffic), it is evident that SAVs can travel at higher speeds with more stable patterns. This stability might be attributed to their capabilities in adjusting headway distances and executing lane changes efficiently. In scenario 7, while the initial speeds are considerably high, similar to the mixed traffic scenarios, the decline in speed is less severe, and the average speed stabilizes at a higher level compared to the baseline and mixed scenarios. This scenario demonstrates the highest overall average speed.

As a conclusion, although integrating SAVs into evacuation scenarios appears to have a positive impact on traffic flow yet, it is also crucial to consider how regular passenger cars and SAVs interact in a mixed traffic environment, which is expected during the initial phase of SAV integration in the future.

Understanding this interaction will be essential for optimizing the benefits of SAV deployment and ensuring smooth traffic dynamics.

Finally, **Figure 4f** depicts the congestion index across various evacuation scenarios. The baseline scenario indicates a significant and steady rise in congestion, peaking sharply towards the end of the evacuation period. This shows severe congestion due to the exclusive use of passenger cars, leading to traffic bottlenecks. Through second to sixth scenarios, congestion index showed similar trends during the initial phase of the evacuation. However, as the percentage of SAVs increased, the congestion index exhibited a noticeable improvement, particularly in the latter part of the evacuation (highlighted in light purple). These scenarios, especially from third to sixth scenario, exhibit lower peaks and more stable congestion levels compared to the baseline, which suggests that higher integration of SAVs contributes to better traffic distribution and reduced congestion. On the other hand, the seventh scenario displays a distinct pattern with a significantly lower congestion index throughout the evacuation period. In this scenario, we have a minimal rise in the congestion index, maintaining a much lower level than the other scenarios. In summary, the simulation results suggest that while a small integration of SAVs has a limited impact on congestion, a higher percentage of SAVs significantly improves traffic distribution and reduces congestion.

Additionally, **Table 4** provides a more microscopic level examination of the percentage change in each metric compared to the baseline scenario. By highlighting the percentage changes for each scenario and metric, it is examined the impacts of SAVs integration by complementing the broader trends observed in the graphical data.

**Table 4.** Evacuation Metrics and Percentage Change Compared to Baseline Scenario Across Different Scenarios

| Metric | | Baseline Scenario $S_1$ | 2nd Scenario $S_2$ | $\Delta(\%)S_2$ | 3rd Scenario $S_3$ | $\Delta(\%)S_3$ | 4th Scenario $S_4$ | $\Delta(\%)S_4$ | 5th Scenario $S_5$ | $\Delta(\%)S_5$ | 6th Scenario $S_6$ | $\Delta(\%)S_6$ | 7th Scenario $S_7$ | $\Delta(\%)S_7$ |
|---|---|---|---|---|---|---|---|---|---|---|---|---|---|---|
| Total Travel Time | $T_{total}$ | $6.3 \times 10^6$ | $6.4 \times 10^6$ | + 1.56% | $5.4 \times 10^6$ | - 14.27% | $4.9 \times 10^6$ | - 21.6% | $4.8 \times 10^6$ | - 23.21% | $4.0 \times 10^6$ | - 35.63% | $4.8 \times 10^4$ | - 99.23% |
| Average Distance | $\bar{d}$ | 24067 | 24205 | + 0.57% | 24731 | + 2.76% | 24448 | + 1.58% | 24844 | + 3.22% | 24388 | + 1.33% | 20602 | - 14.39% |
| Mean Travel Time | $\mu_T$ | 1280 | 1367 | + 6.8% | 1343 | + 4.91% | 1175 | - 8.23% | 1213.88 | - 5.15% | 1102.82 | - 13.83% | 115.34 | - 90.98% |
| Travel Density | ρ | 0.221 | 0.211 | - 4.49% | 0.196 | - 11.65% | 0.184 | - 16.94% | 0.174 | - 21.55% | 0.162 | - 26.94% | 0.02 | - 90.68% |
| Congestion Index | $\xi_C$ | 143.4 | 158.24 | + 10.35% | 156.26 | + 8.98% | 132.48 | - 7.6% | 134.87 | - 5.94% | 119.78 | - 16.46% | 5.19 | - 96.37% |
| Normalized Travel Time | $\eta_T$ | 1280 | 1367 | + 6.79% | 1343 | + 4.91% | 1175 | - 8.23% | 1214 | - 5.15% | 1102.83 | - 13.83% | 115.34 | - 90.98% |
| Average Speed | $\bar{\upsilon}$ | 9 | 8.7 | - 3.24% | 8.9 | - 0.99% | 9.17 | + 1.99% | 9.12 | + 1.47% | 9.44 | + 4.98% | 22.23 | + 147.8% |
| Traffic Volume | $T_{traffic}$ | 5016 | 4625 | - 7.79% | 4300 | - 14.27% | 4089 | - 18.48% | 3850 | - 23.24% | 3745 | - 25.33% | 421 | - 91.6% |
| Travel Efficiency | $\eta_{TE}$ | $3.6 \times 10^{-4}$ | $3.3 \times 10^{-4}$ | - 0.99% | $3.5 \times 10^{-4}$ | - 2.7% | $3.7 \times 10^{-4}$ | + 3.45% | $3.7 \times 10^{-4}$ | + 1.11% | $3.8 \times 10^{-4}$ | + 5.40% | $10^{-3}$ | + 194.8% |

This table presents various evacuation metrics derived from the SUMO, comparing different scenarios against a baseline scenario. Metrics include Total Travel Time (seconds), Average Distance (meters), Mean Travel Time (seconds), Travel Density (vehicles per meter), Congestion Index, Normalized Travel Time (seconds), Average Speed (meters per second), Traffic Volume (vehicles), and Travel Efficiency. To compute the metrics, SUMO's unit conventions were utilized to apply aggregative and central tendency measures. Specifically, Travel efficiency ($\eta_{TE}$) was defined as the ratio of average speed to average distance, $\eta_{TE} = \frac{\bar{\upsilon}}{\bar{d}}$, while the Congestion Index ($\xi_C$) was calculated as the total travel time divided by average speed $\xi_C = \frac{T_{total}}{\bar{\upsilon}}$, evaluated over simulation time intervals. Color coding in the table indicates the performance of each metric compared to the baseline scenario: green signifies improvement, while red denotes deterioration.

**Table 4** reveals significant variations in evacuation metrics across different scenarios compared to baseline scenario: firstly, total travel time shows a noteworthy trend, with substantial decreases observed from the third scenario onwards. Particularly, scenario 7 indicates a dramatic reduction of 99.23% highlighting a significant improvement in overall travel efficiency. As such, it can be suggested that the strategies implemented in the later scenarios effectively reduce the time vehicles spend on the road, improving evacuation performance. When it comes to the average distance, it remains relatively stable across most scenarios, with a slight increase in scenarios 2 to 5. However, scenario 7 deviates with a notable decrease of 14.39% implying that while vehicles generally travel slightly longer distances in earlier scenarios, the final scenario achieves greater efficiency with shorter average distance traveled. For the mean travel time along with normalized travel time, both follow a similar pattern, initially increasing in scenarios 2 and 3, yet showing a marked decrease from scenario 4 onwards. The most significant reduction is observed in scenario 7, with a decrease of 90.98% suggesting that the later scenarios are more effective in reducing the time each vehicle spends traveling, which contributes to more efficient evacuation.

On the other hand, travel density with traffic volume consistently decreases across all scenarios, where scenario 7 experiencing the most substantial decreases of 90.68%, and 91.6%, respectively. This reduction in vehicle density and volume is anticipated to reflect a decrease in the number of vehicles on the road, attributable to enhanced occupancy capacity of SAV. The Congestion Index also mirrors this trend, initially increasing in scenarios 2 and 3 but decreasing significantly from scenario 4 onwards while scenario 7 shows a dramatic decrease of 96.37%, suggesting that later scenarios result in smoother traffic flow and reduced delays.

Average speed demonstrates a slight decrease in scenarios two and three but increases from scenario four onwards, with scenario 7 indicating a notable increase of 147.86%, which is indicative of improved traffic conditions, less congestion, and improved vehicle configurations, allowing vehicles to travel faster and more efficiently. Finally, travel efficiency metric, which measures the ratio of average speed to average distance, shows a minor decrease in scenarios two and three but significantly increases from scenario four onwards, with the most substantial rise of 194.8% in scenario seven, underscoring the overall enhancement in travel conditions.

### 3.2. Evacuation Dynamics under Roadway Closure Constraints

To account for roadway closures in the post-disaster scenarios, a systematic approach was taken by referencing FEMA (Federal Emergency Management Agency) flood hazard zones (FEMA Flood Map Center, 2025) to identify flood-prone road segments, rather than selecting closures randomly. While FEMA flood hazard zones are used for insurance assessment and long-term planning, we applied this data innovatively to identify road segments with heightened vulnerability to flooding during disaster events. The FEMA flood zone interactive map categorizes the target region into three risk levels based on flood hazard classification. The first category, Moderate to Low-Risk Areas (Zone X), represents floodplain areas with a 0.2% annual chance of flooding or less. The second category, High Risk Areas (Special Flood Hazard Areas), includes Zone A, the 100-year floodplain with a 1% annual chance of flooding, and Zone AE, the base floodplain with provided base flood elevations. The third category, High Risk Coastal Areas, includes Zone V and Zone VE, coastal regions with a 1% or greater chance of flooding and additional hazards from storm waves. Based on this classification, we identified two main arterial roads and one urban road within the residential region falling under Zone A. The road closures were strategically implemented to simulate disruptions while making sure alternative routes remained accessible for rerouting. Notably, an on-ramp and off-ramp segment on the Florida Turnpike connecting to the I-75 interstate highway was blocked (**Figure 5a**). This segment is critical as it channels westbound traffic heading north and vice versa, and its closure forces traffic to reroute through alternative highway connections, which leads to significant congestion. Another closure was imposed on an urban road near the residential region, which affects local traffic dynamics (**Figure 5b**). Finally, a segment on I-75 south of the region was blocked (**Figure 5c**), and this is intended to disrupt the primary flow between northbound and southbound direction. To mitigate excessive delays and maintain simulation realism, closures were carefully planned to provide vehicles with alternative routes to their designated exit points. For instance, although segments of I-75 were blocked, parallel side roads remained operational so that vehicles can reroute without overwhelming the simulation or causing excessive delays. Additionally, since scenarios with more than 25,000 vehicles and blocked arterial roads can cause severe traffic congestion and gridlock, we enabled a controlled teleporting mechanism (SUMO, Teleporting, 2025). This approach temporarily removes vehicles during gridlock to resolve the deadlock and reintroduces them later for more smooth traffic yet without relying on excessive teleportation.

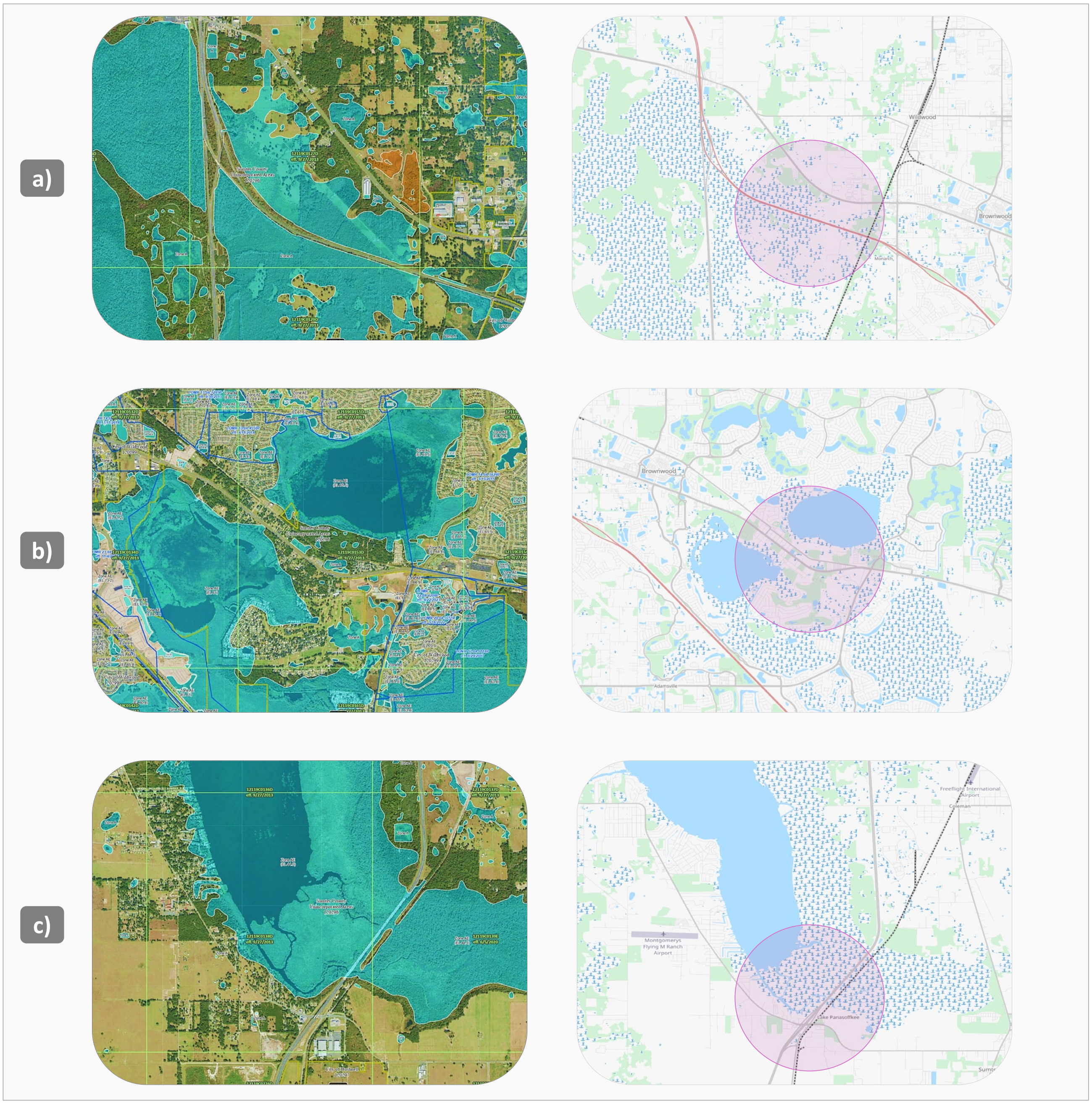

**Figure 5.** FEMA Flood Hazard Zones and Corresponding Road Segment Closures: (a) Displays the on-ramp and off-ramp sections of the Florida Turnpike connecting to I-75 (b) Highlights an urban road segment located in the upper east region of the county (c) Illustrates the I-75 segment in the north-south direction, situated south of the main study region

In configuring demand for post-disaster evacuation scenarios, a slightly different approach was employed compared to pre-disaster cases. As discussed earlier, evacuation departure times for vehicles were established with an upper-level restriction that limits the total evacuation time span within 6-hour window. While this constraint remains consistent, an additional requirement was introduced: the departure times must follow an S-curve pattern, which is a widely accepted approach in evacuation modeling literature. The S-curve captures the characteristic timeline of evacuation demand, where demand initially is low but increases sharply, and peak within the early hours of evacuation. This behavior reflects real-world dynamics, as there is typically a delay between the onset of the evacuation event and the peak in traffic flow due to the latency in evacuation notifications. For instance, a centralized alert with minimal delay can produce a sharper demand curve with a steeper S-shape, while a distributed or delayed notification results in a flatter curve with a more spreader demand over a longer period. For consistency across simulations, we used the same Gaussian distribution parameters to model evacuation start times for all scenarios. However, this process was stochastic and resulted in variations in the S-curve shape across scenarios. For instance, the seventh scenario demonstrated a flatter S-curve distribution compared to others. As shown in **Figure 6**, evacuation dynamics generally exhibit a consistent trend across all scenarios. During the first 30-40 minutes after an evacuation notice, demand remains low, reflecting real-world behavior where individuals take time to prepare or decide on their course of action. This period is followed by a sharp increase in demand, peaking approximately 1.5 hours after the evacuation starts. During this window, a significant proportion of vehicles initiate their departure. After the peak, demand gradually declines most evacuees complete their departures and traffic flow decreases as expected. In **Figure 6**, we have also provided additional statistical metrics, such as probability density and violin plots, to illustrate the distribution of evacuation departure times. These metrics illustrate the variations in stochastic demand configurations while maintaining consistency in the overall observed trends.

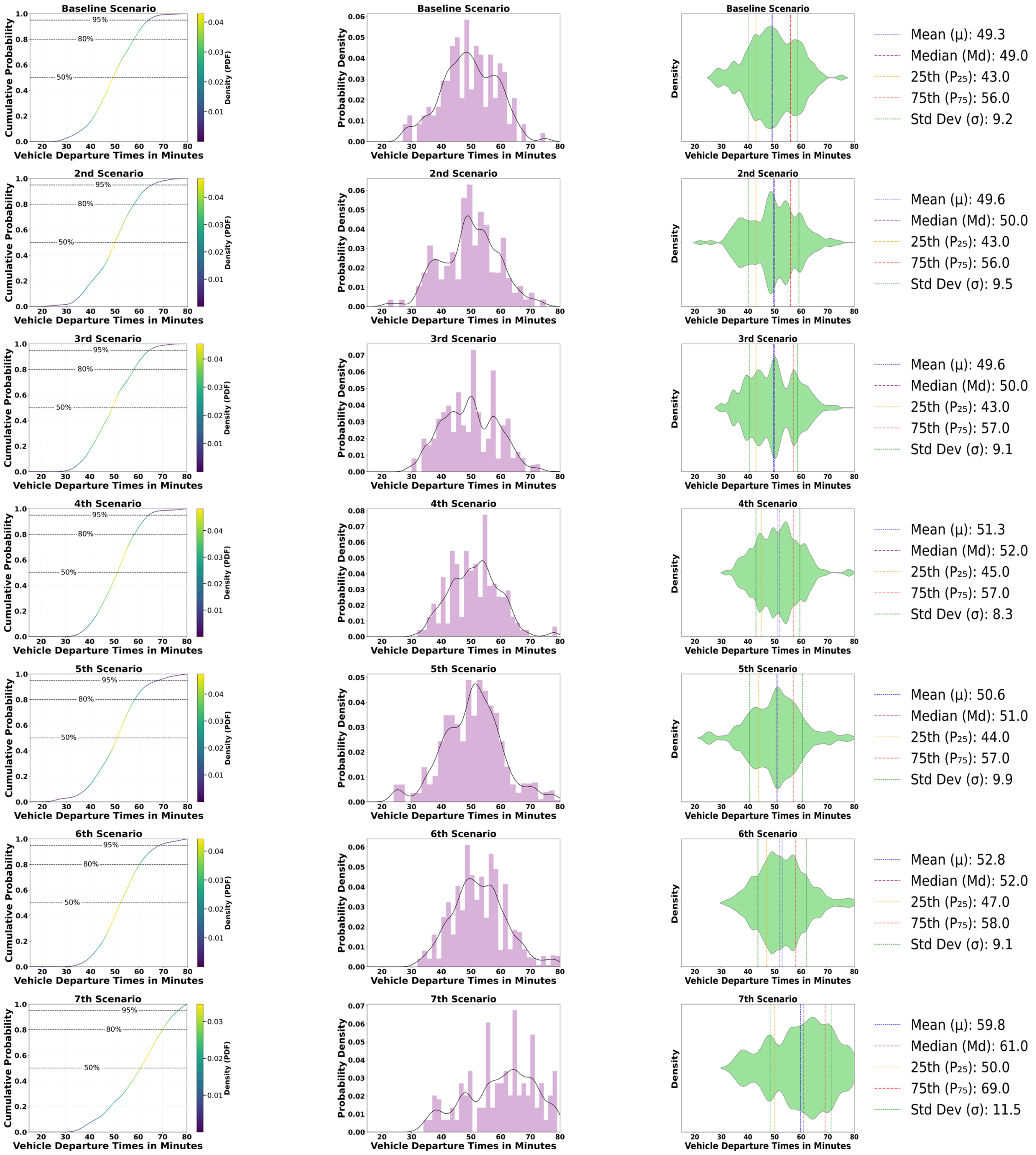

**Figure 6.** Visualization of population demand dynamics using S-curves, probability density, and violin plots. The S-curve illustrates cumulative demand fulfillment over time, the probability density highlights peak demand periods, and the violin plots present statistical metrics of vehicle departure times across scenarios

In the post-disaster evacuation experiments, we analyzed seven different evacuation scenarios to examine their comparative performance under road closure settings **Figure 7**. The findings reveal insights into how road closures influence this type of mass evacuation by considering the advanced vehicle technologies with the purpose of mitigating these impacts. To this end, a key observation is that network clearance times generally show minimal deviations from pre-disaster scenarios for most cases, particularly in Scenarios 2 through 7 (**Figure 7b**). However, as we have seen in **Figure 7a** and **Figure 7b**, vehicles in post-disaster settings require more time to complete evacuations due to road closures and altered traffic dynamics. The baseline scenario, where all evacuees relied on personal vehicles, demonstrated a significant extension in network clearance times. This extension can be attributed to the lack of rerouting capabilities in HDVs, which follows a selfish UE traffic assignment, which eventually leads to severe congestion. Notably, the closure of major arterials, such as Florida Turnpike, diverted traffic to alternative routes, which caused extreme congestion on connected urban and county roads. This effect reiterates the fragility of networks reliant solely on traditional vehicle technologies in handling unexpected disruptions. In contrast, the integration of SAVs resulted in considerable improvements in network clearance times, which is also observed in pre-disaster scenarios, yet this effect is clearer in post-disaster scenarios. The enhanced performance of SAVs can be attributed to their automatic routing capabilities, which allow them to adapt effectively to road closures and traffic congestion. In addition, this observation aligns with the expectation that AV technologies, when strategically deployed, can significantly contribute to evacuation efficiency under constrained network conditions. On the other hand, mean travel times across scenarios did not exhibit clear separation except in Scenarios 2 and 7, which showed comparatively lower values. However, when examining the congestion index (**Figure 7f**) – which is a composite metric that integrates average speed – clearer pattern emerges. Scenarios involving SAVs consistently achieved lower congestion indices, likely due to their ability to maintain higher average speeds with reduced headway distances. In addition, SAVs are less susceptible to getting stuck in severe congestion, allowing them to adhere to their planned evacuation schedules without major arrangements. Following this, an interesting trade-off observed is the relationship between travel distance (**Figure 7c**) and network clearance times. While SAVs exhibited longer average travel distances, this behavior contributed to faster network clearance. This outcome suggests that SAVs prioritize routes that optimize overall evacuation flow, even if it involves covering greater distances. By avoiding congested areas and redistributing traffic loads, SAVs ensure more equitable utilization of network capacity. From the point of autonomous systems design, this trade-off highlights a design consideration achieving a balance between localized efficiency and global network performance. The volume distribution (**Figure 7d**) comes as expected following the strategic departure scheduling. Using an S-curve approach for evacuation departure times resulted in tighter volume distributions, with peak network loads occurring in the mid-phase of evacuation. This contrasts with sparse distributions observed in pre-disaster evacuation experiments and reflects a more realistic representation of evacuation dynamics. Also, the concentration of vehicle volumes during peak periods indicates the importance of advanced traffic signal coordination and adaptive routing algorithms to handle this type of congestion. Speed dynamics across scenarios (**Figure 7e**) provide additional insights into network performance. Average vehicle speeds decreased significantly compared to pre-disaster scenarios due to congestion on major arterials and urban roads. This trend is particularly evident in HDV-dominated scenarios, where vehicles lacked the ability to dynamically navigate around blockages. In contrast, SAVs consistently maintained higher average speeds, although their speed profiles remained below those observed in pre-disaster scenarios.

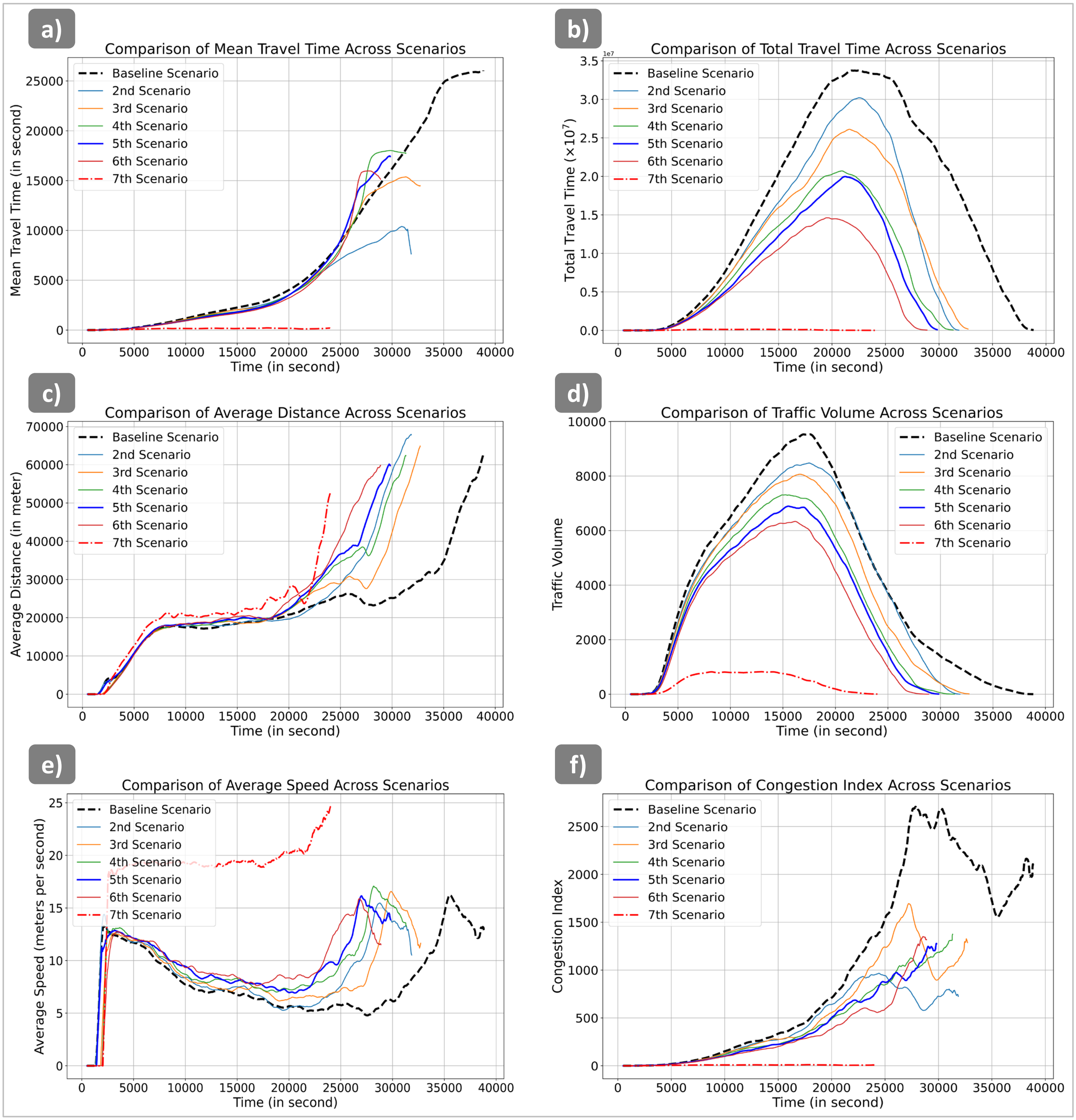

**Figure 7.** Comparison of Post-Disaster Evacuation Metrics Obtained from SUMO Experiments Across Seven Distinct Scenarios

### 3.3. Assessment of SAV-Enhanced Evacuation Metrics through Comparative Analysis with Conventional Buses

In this study, while the integration of SAVs in evacuation scenarios presents numerous opportunities, especially for vulnerable communities, our experiments reveal both promising improvements in evacuation metrics and certain limitations associated with their use. However, it remains essential to assess the model results by testing an alternative mode of transportation during evacuation, such as conventional buses. For this reason, a comparative analysis is conducted by replacing SAVs with conventional buses. This "what-if" scenario aims to investigate whether the improvements observed with SAVs are due to their inherent capabilities or simply coincidental. In these alternative scenarios, conventional buses are set up under identical conditions by making adjustments to reflect their operational limitations. These modifications include but not limited to reducing acceleration, deceleration, and emergency braking capabilities, as well as increasing headway distances to better represent real-world bus dynamics (e.g., *acceleration = 1 m/s², deceleration = 2.5 m/s², emergency deceleration = 5 m/s², sigma = 0.8, minimum gap = 5 meters, tau = 2.0, IcAssertive = 1.3*) (SUMO, Definition of Vehicles, Vehicle Types, and Routes, 2024). It is also important to note that conventional bus capacity (25-person occupancy per bus, with the same occupancy capacity assumed for better comparison of SAV and bus performance), population demand distribution, and SUMO evacuation setup remained identical to those used for the SAV scenarios, except that the buses were not configured with rerouting capabilities, to better mimic real-life conditions. We ran the same seven scenarios configured for SAVs but replaced the SAVs with conventional buses to observe differences in the model results.

Based on the simulation results, **Figure 8** provides a comparative analysis of seven evacuation scenarios to assess both SAV and conventional bus performance across key metrics. As seen in **Figure 8a** and **Figure 8b**, SAVs consistently exhibit lower travel times across all scenarios compared to conventional buses. This can largely be attributed to SAV's advanced route optimization, which allows them to dynamically avoid traffic congestion. Although there are moments, particularly during rerouting phases, where SAVs experience longer travel times, these remain significantly lower overall than those for conventional buses, which tend to become bottlenecked in traffic. On the other hand, when evaluating average travel distances, it is seen that conventional buses perform slightly better in certain scenarios, especially the $3^{rd}$ and $4^{th}$. After the $4^{th}$ scenario, conventional buses demonstrate shorter travel distances. This is likely due to SAVs' tendency to prioritize optimal route selection over minimizing distance, which, while leading to longer distances traveled, enables them to avoid congestion and finally reduce overall evacuation time. This brings a clear trade-off: SAVs may travel further but compensate by minimizing delays, which leads to faster evacuation times. This trade-off becomes particularly clear when analyzing travel speed and congestion levels. **Figure 8e** indicates that in the seventh scenario- where either only SAVs or buses dominate the network- SAVs achieve significantly higher speeds. However, in mixed-vehicle scenarios, in which SAVs or buses are in the network with passenger cars, SAVs are not entirely irresponsive to the effects of traffic congestion, as seen in **Figure 8f**. Although SAVs are generally better at navigating congested environments, they are still susceptible to traffic slowdowns, albeit to a lesser extent than conventional buses.

In conclusion, the results highlight an important consideration for emergency planners: while SAVs shows clear advantages in terms of speed and travel time, they are also subject to increased travel distances in certain moments or scenarios. Thus, if the primary goal is to prioritize the evacuation of vulnerable populations, officials must weigh these trade-offs. SAVs, which are superior in terms of speed and congestion management, may still contribute to congestion without fleet size optimization and under certain conditions. However, their ability to quickly clear evacuees remain a key advantage in the regions where clearance time is a critical factor. Ultimately, we recommend that the choice between SAVs and conventional buses should be determined by regional resources and the region's specific priorities. For instance, if the rapid clearance of an area is the primary objective, SAVs could be a compelling option. Yet, if minimizing travel distances or managing limited infrastructure is a priority, conventional buses may offer a more suitable alternative. The balance between these factors should guide future decision-making in regard to the use of SAVs in emergency scenarios.

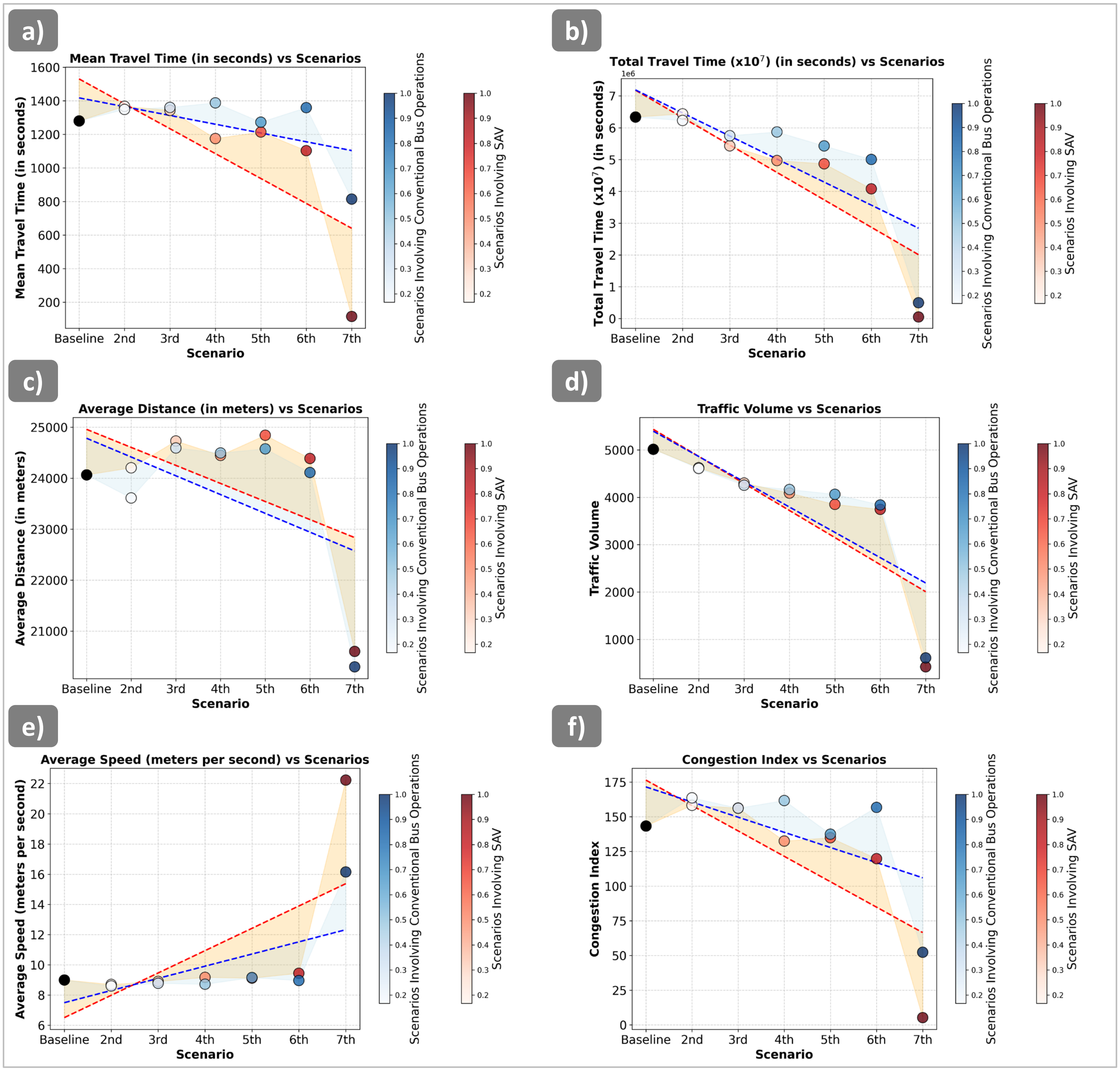

**Figure 8.** The figure illustrates the comparative analysis of key evacuation performance metrics across seven scenarios. The baseline scenario represents an evacuation without SAV or conventional bus integration, while subsequent scenarios progressively increase the proportion of SAVs and conventional buses. For validation, identical simulations were conducted using conventional buses for a direct comparison of performance across both approaches. The analyzed metrics include **Mean Travel Time (a), Total Travel Time (b), Average Distance (c), Traffic Volume (d), Average Speed (e), Congestion Index (f)**. Each subplot provides a comparison of performance between SAVs (indicated by red scatter points and trend lines) and conventional buses (indicated by blue scatter points and trend lines) over the seven scenarios. The baseline, denoted by black marker, serves as a reference point for both vehicle types. The intensity of the scatter points decreases progressively across scenarios. Trend lines are generated via linear regression to identify overall patterns, and shaded regions depict the deviation between observed values and the trend

## 4. DISCUSSION & LIMITATIONS

This study comes with several limitations, which should be addressed in future research. First, as previously mentioned, it was assumed that SAVs are immediately available to the vulnerable populations. However, this is not typically the case in real-world scenarios. While these technologies have the potential to offer fast and efficient services to target communities, passenger waiting times must be studied and incorporated into the relevant cost functions to ensure realistic assessments. Additionally, in the study, SAV bus stops were selected based on the demographic distribution and distance-based criteria of the bus stops (e.g., maintaining a minimum distance between bus stops to avoid congestion). However, it is known that the existing road infrastructure is not fully equipped for these technologies and the placement of SAV bus stops remains under development. In many regions, bus stops will need to be planned and treated differently to accommodate the specific requirements of SAVs. Another significant limitation of the study is that it was assumed there were no fleet size restrictions. In reality, most districts or counties have limited resources, including fleet sizes, which could constrain their ability to evacuate populations effectively. To this end, further research is necessary to determine the optimal fleet size, considering various factors such as demographic, evacuation time span, county's available resources, and road infrastructure. Moreover, while SAVs are equipped with advanced technological features such as LiDAR (Light Detection and Ranging) and cameras, SAVs are simulated using adjusted car-following and lane-changing models within SUMO platform. However, SAVs possess more sophisticated capabilities, which should be integrated into future simulations to achieve more realistic evacuation scenarios. Although the findings of this research provide a solid foundation for understanding the potential benefits and drawbacks of utilizing SAVs in evacuation processes, given the emerging nature of this technology, further research is required to fully grasp its implications and optimize its use for evacuation purposes. Also, a significant limitation of this study stems from the scarcity of reliable and region-specific evacuation trip data. While our simulation aligns with established literature by adapting a Gaussian distribution for evacuation departure times – commonly referred to the S-curve (Li et al., 2013) – validating these dynamics with real-word data remains a challenge. Data sources, such as vehicle counts collected from traffic detectors, are available (Ghorbanzadeh et al., 2021) but often fail to capture the region-specific movement patterns required for accurate validation. Historical datasets, such as the taxi trip data from Hurricane Sandy published by (Donovan & Work, 2016, 2017), offer some insights into population movement during disasters. However, their geographic focus limits their applicability to other regions. Similarly, commercial datasets like those from INRIX, which aggregate geospatial data to track origin-destination flows, diversion routes, and traffic patterns during peak events, could provide another option for calibration. Yet, these datasets are often costly and may not fully represent the unique demographic of all regions. These limitations reiterate the need for region-specific, affordable, and comprehensive evacuation trip data to enhance the accuracy and generalizability of simulation-based studies.

## 5. CONCLUSIONS & FUTURE STUDIES

This study explores the integration of SAVs in rural disaster evacuations through the development of a comprehensive SUMO simulation framework. Based on the findings, it is demonstrated that while SAVs can be useful in improving evacuation efficiency, they come with their own set of pros and cons compared to regular passenger cars. Notably, one of the primary benefits of SAVs is their ability to dynamically reroute,to find the most optimal path to exit routes. By this feature, we configure the shuttles in a way that they are capable of alleviating congestion by distributing traffic more evenly, which, despite longer individual travel times in some scenario configurations, results in a reduced overall evacuation time span. This improvement is particularly crucial for prioritizing vulnerable populations as they are less tolerant to the longer travel times. Moreover, simulation results indicated that higher integration of SAVs considerably improves traffic distribution and reduces congestion, particularly during the latter stages of the evacuation.

In pre-disaster evacuation experiments, scenarios with a higher percentage of SAVs exhibited lower congestion peaks and more stable traffic flow compared to those dominated by regular passenger cars. This suggests that SAVs can optimize route selection more effectively by enhancing overall traffic flow during critical evacuation periods. However, the integration of SAVs is not without its drawbacks. Especially in mixed traffic environments, where both SAVs and passenger cars coexist, it was observed that the average speed tends to decrease. This is expected as SAVs influence traffic dynamics, requiring passenger cars to adjust their trajectories. While we configured SAVs for efficient travel patterns rather than selfish driving, this interaction can lead to slower overall speed. Nevertheless, in scenarios where only SAVs are considered, these vehicles demonstrate higher speeds at more stable travel patterns, because of their advanced capabilities in adjusting headway distances and executing lane changes efficiently. Another observation is the average distance traveled during evacuations. Except for scenarios with complete reliance on SAVs, the average distance traveled in the SAV involved scenarios is slightly more than the baseline scenario in particularly during the later stages of the evacuation, which was attributed to that while SAVs might result in shorter total travel times, they are covering longer distances due to rerouting. Finally, this study underscored the positive impact of integrating SAVs into evacuation scenarios. However, noted that the mixed traffic environment comes with challenges that need to be addressed to fully realize the benefits of SAVs.

The post-disaster evacuation results further validate our findings that while network clearance times across scenarios showed minimal deviations from pre-disaster conditions, road closures significantly impacted evacuation dynamics. In scenarios dominated by HDVs, road closures on major arterials lead to extreme congestion on alternative urban and county roads. In contrast, SAVs demonstrated a clear advantage in post-disaster conditions by adapting to road closures and mitigating the delays based on significantly lower congestion and higher average speeds, even under constraint network conditions. However, mean travel time did not exhibit clear separation across scenarios yet the inclusion of the congestion index highlighted the improved performance of SAVs. Notably, while SAVs tended to travel longer average distances, this behavior facilitated faster network clearance times by prioritizing less congested routes. We also conclude that this trade-off between travel distances and evacuation time highlights the significance of optimizing SAV routing algorithms to balance localized and global efficiency.

According to the simulation results, policymakers should focus on fleet size optimization to balance travel distances and congestion management. Also, combining dynamic rerouting capabilities of SAVs with advanced traffic signal coordination can improve traffic distribution and minimize bottlenecks during peak evacuation periods. Addressing the interactions between SAVs and HDVs in mixed traffic environments is critical and developing cooperative traffic management strategies can ensure smoother integration. Finally, SAV deployment strategies should prioritize vulnerable groups to minimize their in-vehicle travel times and improve evacuation outcomes.

In addition, there are several key areas that future research should focus on to further enhance the efficacy of SAVs in disaster evacuation scenarios:

1. Consideration of diverse natural disaster scenarios and road capacity impacts: Future studies should explore a variety of natural disaster scenarios, such as hurricanes, fires (e.g., the recent Los Angeles wildfires), and floods, which could potentially affect road capacity both spatially and temporally. Understanding how different types of disasters impact evacuation routes and available road infrastructure will enable more flexible and robust planning. Furthermore, this would include simulation frameworks to account for real-time changes in road conditions which may offer a more realistic portrayal of the challenges in evacuation scenarios.
2. Enhancing SAV integration in mixed traffic scenarios: Although this study utilizes simulation-based methods, such approaches may not fully capture the potential of SAVs, particularly in mixed traffic environments. To address this limitation, it is recommended to consider solving a Multiclass Traffic Assignment Problem rather than relying solely on analytical solutions. For example, previous research (Bamdad Mehrabani et al., 2023) has proposed an open-source framework for multiclass, simulation-based traffic assignment in mixed traffic scenarios involving AVs and HDVs. Their model assumes that AVs follow a system-optimal routing with dynamic rerouting, while HDVs adhere to UE traffic assignment. Adapting such a framework to evacuation scenarios could provide deeper insights into the impact of SAVs on mixed traffic flow dynamics. ,

3. Incorporation of advanced optimization algorithms for improved route assignment: To further enhance the effectiveness of SAVs in disaster scenarios, integrating advanced optimization algorithms into the routing process and test and make comparisons among them could yield significant improvements in evacuation efficiency. These algorithms, namely machine learning-based optimization or multi-objective optimization techniques, could help SAV better navigate complex evacuation routes, respond to real-time traffic data, and dynamically reroute based on changing conditions to make the evacuation process more adaptable and efficient.

## AVAILABILITY OF DATA AND MATERIALS

The datasets used and/or analyzed during the current study are available from the corresponding author on reasonable request.

## COMPETING INTEREST

The authors declare that they have no competing interests.

## FUNDING

This research was funded by the Rural, Equitable and Accessible Transportation (REAT) Center, a Tier-1 University Transportation Center (UTC) funded by the United States Department of Transportation (USDOT), through the agreement number 69A3552348321. It is also funded by National Science Foundation Award # 2400153. The contents of this paper reflect the views of the authors, who are responsible for the facts and the accuracy of the information presented herein. The U.S. Government assumes no liability for the contents or use thereof.

## AUTHOR CONTRIBUTIONS

Alican Sevim: Methodology, Investigation, Formal analysis, Visualization, Writing - original draft. Qianwen Guo: Conceptualization, Validation, Writing - review & editing, Supervision. Eren Ozguven: Writing - review & editing, Supervision

## ACKNOWLEDGMENTS

Not applicable